\documentstyle[twocolumn,aps,epsfig]{revtex}
\makeatletter
\def\thepage{3-\@arabic\c@page}
\def\@pnumwidth{2em}

\makeatother

\newcommand {\bean}  {\begin{eqnarray*}}
\newcommand {\eean}  {\end{eqnarray*}}

\def\Rset {{\rm I \kern-.2em R}} 

\newcommand {\ds}   {\displaystyle}

\newcommand {\bce}  {\begin{center}}
\newcommand {\ece}  {\end{center}}
\newcommand {\be}   {\begin{equation}}
\newcommand {\ba}   {\begin{array}}
\newcommand {\bea}  {\begin{eqnarray}}
\newcommand {\bfi}  {\begin{figure}}
\newcommand {\ee}   {\end{equation}} 
\newcommand {\ea}   {\end{array}}
\newcommand {\eea}  {\end{eqnarray}}
\newcommand {\efi}  {\end{figure}}

\newcommand {\noi}  {\noindent}

\newcommand {\UNIV}   {Universit\`a }

\def\Rset {{\rm I \kern-.2em R}} 
\def\mathbbH {{\rm I \kern-.2em H}} 
\def\mathbbC {{\rm I \kern-.6em C}} 

\begin{document}
\title{Homogeneous isotropic turbulence in dilute polymers: 
scale by scale budget 
\vskip 1pc
%
E\@. De Angelis
\thanks{Dip. Mecc. Aeron., \UNIV di Roma {\em La Sapienza},
        via Eudossiana 18, 00184, Roma, Italy.},
C\@.M\@. Casciola$^*$,
R\@. Benzi
\thanks{Dipartimento di Fisica e INFM, \UNIV di Roma {\em{Tor Vergata}},
Via della Ricerca scientifica 1, 00133 Roma, {{Italy}}},
\& 
R\@. Piva$^*$ }
\maketitle
\makeatletter
\global\@specialpagefalse
\let\@evenhead\@oddhead
\def\@oddfoot{\reset@font\rm\hfill \thepage\hfill
\ifnum\c@page=1
  \llap{\protect\copyright{} 1996
  American Institute of Physics}%
\fi
} \let\@evenfoot\@oddfoot
\makeatother
\section*{Abstract}
The turbulent energy cascade in dilute polymers solution is addressed here by considering 
a direct numerical simulation of homogeneous isotropic turbulence of a FENE-P fluid in 
a triply periodic box. On the basis of the DNS data, a scale by scale analysis is 
provided by using the proper extension to visco-elastic fluids of the Karman-Howarth 
equation for the velocity. For the microstructure, an equation, analogous to the Yaglom 
equation for scalars, is proposed for the free-energy density associated to the elastic 
behavior of the material.
Two mechanisms of energy removal from the scale of the forcing are identified, namely 
the classical non-linear transfer term of the standard Navier-Stokes equations and
the coupling between macroscopic velocity and microstructure. The latter, on average, 
drains kinetic energy to feed the dynamics of the microstructure.
The cross-over scale between the two corresponding energy fluxes is identified, with 
the flux associated with the microstructure dominating at small separations to 
become sub-leading above the cross-over scale, which is the equivalent of the  
elastic limit scale defined by De~Gennes-Tabor on the basis of phenomenological assumptions.

\section{Introduction}

Turbulence in dilute polymers solutions is still an open issue, despite 
the growing interest on the subject.
Most of the efforts have been addressed to the comprehension of practical 
aspects, such as drag reduction, related to the modification of turbulence 
by long chain polymers in wall bounded flows, \cite{Lumley}. 
However, in most cases, the
attempts have proven inconclusive, suggesting that, even from the point of 
view of applications, a more fundamental approach is required.
Under this respect, basic studies should focus on simpler flow 
conditions where turbulence-polymers interaction occurs.  

A possible choice is to exploit homogeneity and isotropy to simplify 
the statistical treatment of the data and to make use of the available, 
relatively well established, rheological models for dilute polymers, 
see \cite{Bird} for a comprehensive review.

In this context, a recent experimental investigation on decaying grid turbulence 
of dilute polymers and surfactants  \cite{Sreeni2} shows, for the polymers, a 
substantial alteration  of the small scales, though still in presence of a substantial 
degree of anisotropy, ascribed to the grid shear layers.

In this and other related experiments, see e.g.  \cite{Mcomb} 
and \cite{Schwartz}, the measurements were mostly aimed at the velocity,
but also at the pressure field see e.g. \cite{Cadot}
for a recent example.
Actually, as one can imagine, the experimental analysis of turbulent flows 
of polymeric liquids is particularly difficult, and hardly one can proceed 
beyond the global characterization of the flow and the analysis of certain
aspects of the macroscopic  field, see \cite{Tiederman}, 
\cite{warholic} as additional examples in the more complex configuration
of a channel flow.
This makes the actual mechanism
of interaction between polymers and turbulence particularly obscure, and
leads to consider numerical simulations as a viable tool to address the 
problem.  For triply-periodic boundary conditions, in particular, fast
and accurate spectral methods can easily be developed to achieve the 
direct numerical simulation of a strictly homogeneous isotropic flow, 
once a reasonably accurate and sufficiently simple rheological model 
has been selected.
Among these, the so-called FENE-P model \cite{Bird2} is possibly the best
compromise between accurate representation of polymers dynamics and
minimal computational complexity, see \cite{Pilg}.
In fact, recent results show that this model is able to reproduce
the drag reducing behavior of dilute polymers in wall bounded flows,
see \cite{Beris}, \cite{DA1}, \cite{DA2}.

As often in rheology, one can easily get stuck in a discussion about
the proper model for a specific application. 
We take here the model for granted and try to grasp the mechanisms
by which it affects the turbulence. To do this we take 
freedom to chose parameter values which may not correspond too well to 
experimental conditions, but which may help revealing the underlying physics.

In summary, our model for the solution considers an ensemble of elastic dumbbells
attached to each material point of the continuum. The dumbbells are stretched by 
the flow due to friction and react through a non-linear elastic spring.
The probability density function of the end-to-end vector of the dumbbells is 
described through a second order tensor, the conformation tensor, assumed to represent
the covariance matrix of the end-to-end vectors. It obeys a transport equation
which accounts for material stretching and for the elastic reaction.
The force exerted on the continuum by the dumbells contributes an 
extra-stress term in the macroscopic momentum balance.
On average, the work done by the extra-stress against the macroscopic velocity field
amounts to a draining of energy from the macroscopic field. The corresponding power
feeds the dynamics of the microstructure, is partially accumulated as free-energy
of the ensemble and is eventually dissipated, see \cite{mariano}.

In terms of turbulence dynamics, several basic questions arise.
First of all, one should understand if 
the nature of this kind of viscoelastic  turbulence is substantially
similar, or, on the contrary, essentially different from standard Newtonian 
turbulence. 
According to the common view, a turbulent flow is described as a superposition
of different scales of motion. In homogeneous isotropic turbulence of ordinary fluids 
the interaction of the different scales originates a flux of energy from the large towards 
the small scale, with energy injected by an external forcing at large scales and 
dissipated by viscosity at small scales, the so-called direct, or forward, energy cascade
(see e.g. the book by Frisch \cite{frisch}).
The presence of the microstructure is expected to alter this process, since it corresponds to
an alternative path the energy flux may take to achieve the eventual dissipation.
If this is the case, one may wonder if the alteration is localized at certain spatial
scales or if the entire spectrum is affected by the energy draining of the polymers.
This poses the additional issue of determining a possible scale which delimits the 
range of scales where the presence of the polymers is effective from those where they
are substantially passive, see \cite{De Gennes}.
This scale may be close to the classical 
Kolmogorov dissipation scale, but it may even be substantially larger, and 
it is not even
clear if the effect of polymers may be reduced to an additional dissipation,
\cite{Lumley2}, or if the role of elastic energy is crucial, \cite{De Gennes},
see also \cite{Sreeni}.

All these issues can be addressed systematically by analyzing the data of a DNS of
homogeneous isotropic turbulence once proper diagnostic tools are available.
To this purpose we present and discuss a scale by scale budget based on the extension
to viscoelastic fluids of the classical Karman-Howarth \cite{frisch} equation of 
ordinary turbulence, see e.g. \cite{hinze}, \cite{Oberlack} and \cite{casciola} for 
similar extensions in the context of shear flows.
By this equation we can address the effect of polymers on the
macroscopic kinetic field. To consider the fluctuations in the polymers
we would need to extend the classical Yaglom equation for a scalar, see \cite{Monin}, 
to a second order active tensor.
In a more straightforward way instead we consider, in the proper thermodynamic framework, 
the Yaglom equation for the free-energy of the polymers. 

\section{The evolution equations for dilute polymer solutions}

As discussed in the introduction, the momentum balance for a dilute solution
of long chain polymers in otherwise Newtonian, incompressible solvent is described 
by a slightly modified form of the Navier-Stokes equation, namely

\bea
\label{NS}
\frac{D u_i}{Dt} & = & \frac{\partial p}{\partial x_i} \, + \nu \, 
\frac{\partial^2 u_i}{\partial x_j \partial x_j} \, + \,  \frac{\partial T_{i j}}{\partial x_j} \, + 
f_i \, ,
\eea

\noi where $D /Dt = \partial/ \partial t \, + \, u_k \, \partial /\partial x_k$ is the usual
substantial derivative, $f_i$ the external forcing, $p$ is the pressure normalized by the 
constant density of the solution, 
$\nu$ is the kinematic viscosity of the solvent and $T_{i j}$ is the additional contribution 
to the stress tensor - extra-stress - due to the polymers chains. 
The macroscopic velocity field $u_i$ is solenoidal and the constitutive relation
for the extra-stress is

\bea
\label{ex-stress}
T_{i j} & = & \nu_p/\tau  \, (\, f(R_{kk}; \rho_{m}, \rho_0) R_{i j}/\rho_0^2  - \delta_{i j}\, ) \, ,
\eea

\noi where $\nu_p$ is a constant typically of the order of a small but finite fraction of $\nu$.
The conformation tensor, $R_{ij}$, is a second order tensor taken to characterize the 
statistical behavior of the population of polymers chains attached to a given point, and
$f(R_{kk}; \rho_{m}, \rho_0)$ is a dimensionless nonlinear spring coefficient specified 
as $f = (\rho_m^2 - \rho_0^2)/(\rho_m^2 - R_{kk})$,
with $\rho_0$ the equilibrium  length of the chains and $\rho_{m}$ their maximum allowed
length, as standard in the context of the finite extensibility nonlinear elastic model 
(FENE, see \cite{Bird}).
The dynamics of the population of chains is described by an evolution equation for the
conformation tensor,

\bea
\label{EQ_CT}
\frac{D R_{i j}}{D t} & = & 
\frac{\partial u_i}{\partial x_k} R_{k j} \, + \, R_{i k} \frac{\partial u_k}{\partial x_j} \, 
- \frac{1}{\tau}(f R_{i j} - \rho_0^2 \delta_{i j}) \, ,
\eea

\noi where $\tau$ is the principal relaxation time of the chains.
The system (\ref{NS}, \ref{ex-stress}, \ref{EQ_CT}) completed with the continuity equation,

\bea
\label{CE}
\frac{\partial u_i}{\partial x_i} & = &  0 \, ,
\eea

\noi for the macroscopic velocity forms the FENE-P model for dilute polymers solutions, 
see e.g. \cite{Bird2}.

\noi Phenomenologically, stretching of the chains by the flow, described by the terms involving
the velocity gradient in (\ref{EQ_CT}), is counteracted by an elastic reaction, the term in brackets,
while the polymers are transported and re-oriented by the flow. The reaction of the polymers 
alters the macroscopic force  balance, as seen in eq.~(\ref{NS}), where the divergence of the
extra-stress enters as an additional source or sink of momentum.

The extra-stress is able to make work against the velocity as follows from
the balance of kinetic energy 

\bea
\label{KE_b}
\frac{D (u^2/2)}{D t} = \frac{\partial \,(p \, \delta_{i j} + \Sigma_{i j}) \,u_i}{\partial x_j} \, + 
\, \frac{\partial \, ( T_{i j} \, u_i)}{\partial x_j} -  
\, \epsilon_N + \, S \, , 
\eea

\noi where $\Sigma_{i j} = 2 \nu e_{i j}$ is the standard viscous component of the stress tensor and
$\epsilon_N = 2 \nu e_{i j} e_{i j}$ is the Newtonian component of the dissipation, with
$e_{i j}$ the macroscopic velocity deformation tensor. With respect to the kinetic energy balance
of the standard Navier-Stokes equation, two additional terms arise here, to describe the interaction
of the macroscopic field with the microstructure. The one in divergence form contributes to the 
spatial redistribution of kinetic energy. The other,

\bea
\label{SP0}
S = T_{i j} \frac{\partial u_i}{\partial x_j} \, ,
\eea

\noi called hereafter the stress-power, represent energy per unit time that the microstructure can drain, 
or, in principle, even release to the macroscopic kinetic field.
The presence of this term is physically due to the fact that the polymers chains can both dissipate, via
friction with the solvent, or store energy, due to their elasticity.
The correct thermodynamical setting refers to the free-energy $a$ of the polymers ensemble.
Given the expression, see e.g. \cite{Bird},

\bea
\label{a-def}
a = 
-\frac{1}{2} \frac {\nu_p}{\tau} 
\left[ ( \rho^2_{max}/\rho_0^2-1) 
\log\left( \frac{\rho^2_{max} - {R}_{kk}}{{\rho^2_{max} -\rho_0^2}} \right) \right.  \nonumber \\
 \\
\left. \qquad \qquad +1/3 \,  \log(\det{ R/\rho_0^2})\right] \, ,  \nonumber 
\eea

\noi it follows directly from the evolution equation (\ref{EQ_CT}) that the energy balance for the
polymers is

\bea
\label{F_En}
\frac{D a}{D t} & = & S - \epsilon_P \, ,
\eea

\noi where 

\bea
\epsilon_P = 
\frac {1}{2} \frac {\nu_p}{\tau^2} f \, 
\left[ (f \, R)_{kk}  +  (f \, R)^{-1}_{kk} - 6\right] 
\eea

\noi a positive definite quantity, is the rate of energy dissipation 
associated with the polymers.  
Adding eq.~(\ref{KE_b}) and (\ref{F_En}) gives the balance of the total free-energy of the system,
$E = u^2/2+a$, as

\bea
\label{C_GE}
\frac{D E}{D t} & = & f_i u_i -\epsilon_T + \frac{\partial J_k}{\partial x_k}
\eea

\noi where $J_i$ is the spatial flux of total free-energy, whose expression 
follows immediately by inspection of the kinetic energy balance.
Note that $S$ drops from the global balance since it corresponds to a conservative exchange of 
energy between the two constituents, namely the macroscopic kinetic energy and the polymers 
free-energy.

\section{The Karman-Howarth equation}

In a statistically stationary turbulent system, the external forcing $f_i$ 
provides the energy to maintain the turbulent fluctuations which 
otherwise would be damped by the dissipation. For the present case both the macroscopic
velocity and the microstructure exhibits turbulent fluctuations.
Characteristic feature of the system is that, while the external forcing
acts only on the macroscopic field, both the macroscopic flow  and the
microstructure contribute to the dissipation of the injected mechanical
power. This is possible since the extra-stress captures part of the
available mechanical energy to feed the fluctuations of the 
microstructure and the associated dissipation.
In order to try to disentangle these complex processes, the simplest
setting is provided by homogeneous and isotropic turbulence once
suitable balance equation are available for the fluctuations of 
both kinds.

Purpose of the present section is to present the equation appropriate
to address the scale by scale budget of the macroscopic velocity 
fluctuations.
This is the extention to dilute polymer solutions of the well known
Karman-Howarth equation of homogeneous isotropic turbulence 
for Newtonian fluids. 
As discussed in the previous section, where the FENE-P model has been briefly recalled,
the solenoidal velocity field, $u_i$, obeys a modified form of
the Navier-Stokes equation, where the divergence of the extra-stress
$T$, denoted here by  

\bea
\label{g_def}
g_i  & = & \frac{\partial T_{ij} }{\partial x_j} \, ,
\eea

\noi appears as an additional source of momentum.
 
Starting with the modified Navier-Stokes equation, the equation for
the correlation tensor $C_{i,j} = <u_i\, u'_j>$, where unprimed and primed 
variables are evaluated at $x_i$ and at $x'_i$, respectively, with $x'_i$
the displacement of $x_i$ by the separation vector $r_i$,
easily follows from the standard procedure, see e.g. \cite{hinze}, \cite{Monin}.
The trace of this equation reads

\begin{eqnarray}
\label{C_ii}
\frac{\partial C_{i,i}}{\partial t} + \frac{\partial}{\partial r_k}
<u_i u^\prime_i u^\prime_k - u_i u^\prime_i u_k>  =  \qquad \qquad \nonumber \\
2 \nu \frac{\partial^2 C_{i,i}}{\partial r_k \partial r_k } \, + \, 
<g'_i u_i + g_i u'_i> \, + \, 
<f'_i u_i + f_i u'_i> \, , \nonumber \\
\end{eqnarray}

\noi where $f_i$ is here specified as a random stirring force acting on 
the large scales.
To obtain the above equation we have used standard properties
of homogeneous incompressible turbulence, which allow to drop the pressure
term and to recast spatial derivatives with derivatives with respect to 
the separation, see e.g. \cite{frisch}.

Equation~(\ref{C_ii}) can be re-expressed in terms of velocity increments,
    $ \ds  \delta V_i= u_i(x'_r) - u_i(x_r) $, 
$\; \delta V^2=\delta V_i \delta V_i$, once recalled that 
$ C_{i,i}=<u_i u_i> - 1/2 <\delta V^2> $,
and that, by standard manipulations, we have 
\bea
\label{dv2_dv_k}
<\delta V^2 \delta V_k> = -2 <u_i u^\prime_i u^\prime_k-u_i u^\prime_i u_k> + \nonumber \\
\qquad \qquad \qquad \qquad <u_i u_i u^\prime_k>-<u^\prime_i u^\prime_i u_k> \, .
\eea
Hence, for a solenoidal velocity field, the equation in terms of velocity increments 
follows as 
 
\bea
\label{k_h_diff}
\frac{\partial}{\partial r_k}<\delta V^2 \delta V_k
+2  T^*_{k i} \delta V_i >  = 
-4 <f_i u_i> \, + \,  \nonumber \\
 2  <\delta V_i \delta f_i> \, + \, 
2 \nu \frac{\partial^2}{\partial r_k \partial r_k} <\delta V^2>. \nonumber \\
\eea
\noi In deriving eq.~(\ref{k_h_diff}) a statistically stationary field has been assumed 
and the term $<f'_i u_i + f_i u'_i>$  has been rearranged in the form 
$2 <f_i u_i> - <\delta V_i \delta f_i>$ to extract explicitly the power injected by the 
external forcing. 
Finally, the integration by parts 
\bea
\label{T_star}
<g'_i u_i + g_i u'_i> = 
\frac{\partial }{\partial r_k} ( T'_{k i} u_i - T_{k i} u'_i ) = 
- \frac{\partial }{\partial r_k} ( T^*_{k i} \delta u_i)  \, , \nonumber \\
\eea

\noi where $T^*_{k i} =  T'_{k i} + T_{k i}$, 
leads to eq.~(\ref{k_h_diff}).

Equation~(\ref{k_h_diff}) can be integrated in $r$-space over a ball $B_r$ of radius $r$
(see e.g. \cite{casciola} for a similar approach) to yield

\begin{eqnarray}
\label{k_h_sphere}
\frac{1}{4 \pi r^2}
\oint_{\partial B_r} <\delta V^2 \delta V_k + 2 \,
 T^*_{k i} \, \delta V_i> n_k \, d S_r   =  
  -4/3 \, \bar{\epsilon}_T \, r  
\nonumber  \\  
+  \frac{\nu }{2 \pi r^2} \,
\frac{d}{dr} \,  \oint_{\partial B_r}  <\delta V^2 > \quad d S_r \, 
\nonumber  \\  
+ \frac{1}{2 \pi r^2}
\int_{B_r}  <\delta V_i \delta f_i> \quad d V_r  \, ,
\nonumber \\
\end{eqnarray}
\noi where the global 
energy balance (\ref{C_GE}), which for a stationary homogeneous field reads

\bea
\label{kin_bal_av}
<f_i u_i> & = & {\bar \epsilon}_T \, ,
\eea

\noi has been accounted for.

The Karman-Howarth equation (\ref{k_h_sphere}) highlights the main difference between 
Newtonian and viscoelastic turbulence. For either case, in the range of scales where 
the direct effect of viscosity and the correlation between velocity and forcing increments
are both negligible, the energy flux occurring through scale $r$ equals the total 
dissipation. For Newtonian fluids, all the dissipation is provided by the viscosity
and the flux through scale $r$ only occurs due to the classical nonlinearity associated to
the advection.
For viscoelastic fluids, an additional dissipative process takes place in the polymers,
and, consistently, the flux through scale $r$ also presents a new component in 
addition to that of the standard Navier-Stokes dynamics.
In fact, the additional flux term describes the amount of energy intercepted by the 
microstructure.  
Though the classic component is expected to be associated to a forward energy cascade,
i.e. the corresponding flux enters the ball of radius $r$, concerning the 
viscoelastic contribution we can, at the moment, only conjecture that energy
intercepted at scale $r$ is passed forward to the microstructure at smaller scale,
consistently with the idea of polymer dissipation as a small scale process.
However, since, to our knowledge, the process has never been investigated before, 
we have to examine and discuss the numerical simulation before drawing any conclusion.

\section{The Yaglom equation for the free-energy}

The Karman-Howarth equation for dilute polymer solution may be used
to analyze the scale by scale budget for the velocity field.
Its Fourier transform actually corresponds to the budget of turbulent
kinetic energy in the different modes of the spectrum.
To address the complete dynamics of dilute polymers solution we need
a similar equation concerning the microstructure.
Under this respect, in section~1 we have shown 
that a free-energy $a$ can be associated to the ensemble of polymers
attached to a given point in space and that, locally, the energy balance 
for the microstructure is provided by equation~(\ref{F_En}).

The free-energy $a$ is a measure of the average elongation of the polymers 
as given by the trace of the conformation tensor $R_{i j}$. 
In fact one can derive an equation for the scale by scale budget 
of the trace though we prefer to discuss the Yaglom equation for $a$, whose
meaning in terms of energetics of the system is more definite.
To this purpose, let us denote by 
 
\bea
\label{psi}
\psi = S - \epsilon_p
\eea

\noi the instantaneous excess power, i.e. the amount of power transfered to 
the polymers in excess to the dissipation they originate. 
Strictly speaking this is an improper naming, since on the average the excess power is zero.
Nonetheless, it measures the amount of energy stored per unit time in the polymers. 
From (\ref{F_En}) we can derive the equation for the correlation $<a a'>$,

\bea
\label{a-corr}
\frac{\partial <a a'>}{\partial t} \, + \, \frac{\partial }{\partial r_k} <u'_k a' a - u_k a a'> & = & 
\nonumber \\
<a' \psi + a \psi'> \, ,
\eea

\noi which, considering that

\[ 
<a \psi' + a' \psi> = 2 <a \psi> - <\delta a \delta \psi> \, ,
\]

\noi and noting that, for a stationary state ($ <a \, {D a}/{D t}> = 0 $),

\bea
\label{zero_diss}
<a \psi> = 0 \, ,
\eea

\noi can be recast in the form

\bea
\label{Y-eq}
\frac{\partial }{\partial r_k} <\delta u_k \delta a^2> = 2 <\delta a \delta \psi> \, .
\eea

\noi Integration over the sphere of radius $r$ then yields

\bea
\label{Y_sphere}
\frac{1}{4 \pi r^2 }\oint_{\partial B_r} <\delta a^2 \delta V_k> n_k d S_r  & = &  \nonumber \\
\frac{2}{4 \pi r^2 }\int_{ B_r} <\delta a \delta \psi> d V_r   \, ,
\eea

\noi which, by using isotropy reduces to

\bea
\label{Y_parallel}
<\delta a^2 \delta V_\parallel>  & = & 2 \int_0^r <\delta a \delta \psi > r^2 dr \, ,
\eea

\noi with $\delta V_\parallel = \delta V_i \, r_i/r$ the longitudinal velocity increment.

Equation (\ref{Y_parallel}) is the appropriate form of the Yaglom equation for the polymers.
It states the equivalence between the flux of free energy through the different spatial scales,
as described by $<\delta a^2 \delta V_\parallel>$ and the integral of the correlation between 
free energy fluctuations, $\delta a$, and fluctuation of the excess of power, $\delta \psi$.
Classically for scalars the Yaglom equation provides a term linear in the separation
associated to the average scalar dissipation. In the present case, such term is zero
as consequence of eq.~(\ref{zero_diss}).

\section{Global features of the flow and DNS}

The system of equations (\ref{NS},\ref{EQ_CT},\ref{CE}), corresponding to 
$9$ scalar equation, $3$ velocity components plus $6$ components of the 
conformation tensor, has been integrated using a Fourier $\times$ Fourier 
$\times$ Fourier spectral method for spatial 
discretization and a four-stages/third order Runge-Kutta method for time
advancement with all the non-linear terms fully de-aliased by the 
three-halves rule.
An exactly solenoidal velocity is achieved by the standard projection method.
The side of the cubic computational box is $2 \pi$ and $64$ modes are used in 
each direction.
A fully developed field of Newtonian turbulence has been used
as initial condition for the viscoelastic calculation and
the random forcing is applied to the first shell of wave-vectors, with
constant amplitude and uniformly distributed phases.
The random phase extracted at each step is used subject to the condition 
of positive energy injection, otherwise it is changed by $\pi$.

The nominal Reynolds number of the simulation can be constructed from the 
intensity of the forcing, $\hat f_0$, the length of the box, $l_x$, and the 
zero shear kinematic viscosity, $\nu_T = \nu + \nu_P$, as 
$Re = (l_x^3 \hat f_0)^{(1/2)}/\nu_T = 676$.
The parameters concerning the polymers are $\eta_p = \nu_p/\nu = .1$, see 
e.g. \cite{Sreeni2}, 
$\rho^2_m/\rho^2_0 = 1000$ and $De = \tau (\hat f_0/l_x)^{1/2} = .38$.
When comparing viscoelastic and Newtonian simulations, we keep constant the
nominal Reynolds number, i.e. for the Newtonian case we use the same forcing 
intensity and a value for the viscosity equal to the value of $\nu_T$ for 
the viscoelastic fluid.

After the transient is elapsed, statistically stationary conditions are achieved.
The general impression on the field may be gained by considering figure~\ref{F1},
where instantaneous vortical structures are shown for the Newtonian and the viscoelastic
simulation. The eduction criterion for vortical structures is based on the discriminant
of the velocity gradient tensor, one of the now classical approaches to extract vortical
structures from a complex field, see e.g. \cite{chong} for details.
Algebraically, the idea is to identify with vortical structures those regions where
a couple of eigenvalues of the velocity gradient are complex conjugate, implying
a locally helicoidal relative motion of nearby particles.
In principle, other criteria could be employed and different field quantities could be 
visualized. In any case, the inspection of the instantaneous fields always leads to the
conclusion that, with respect to Newtonian, much less but wider structures are present 
in the viscoelastic case.

\bfi[t!]
   \centerline{
   \epsfig{figure=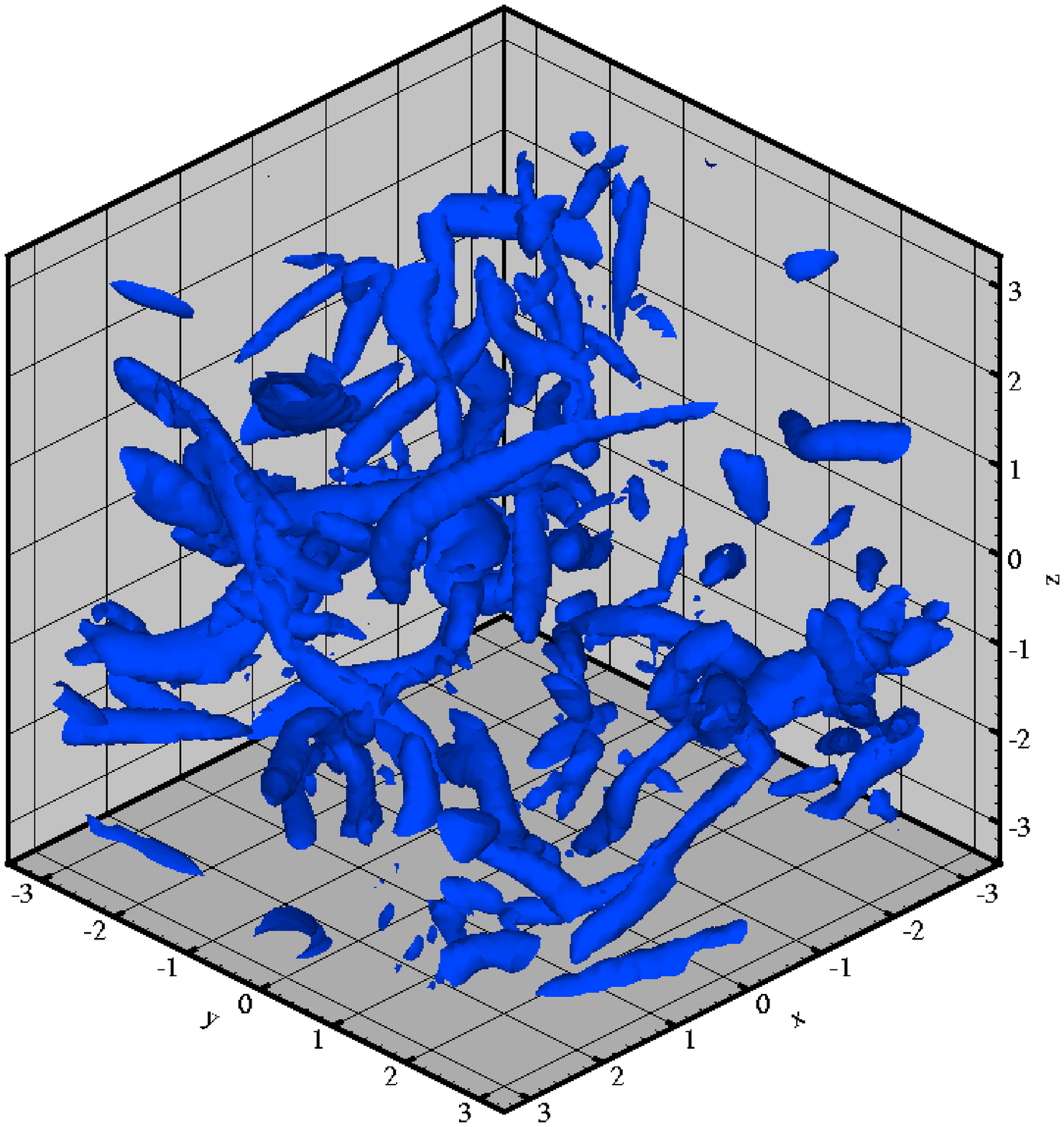,width=7.cm} }
   \centerline{
   \epsfig{figure=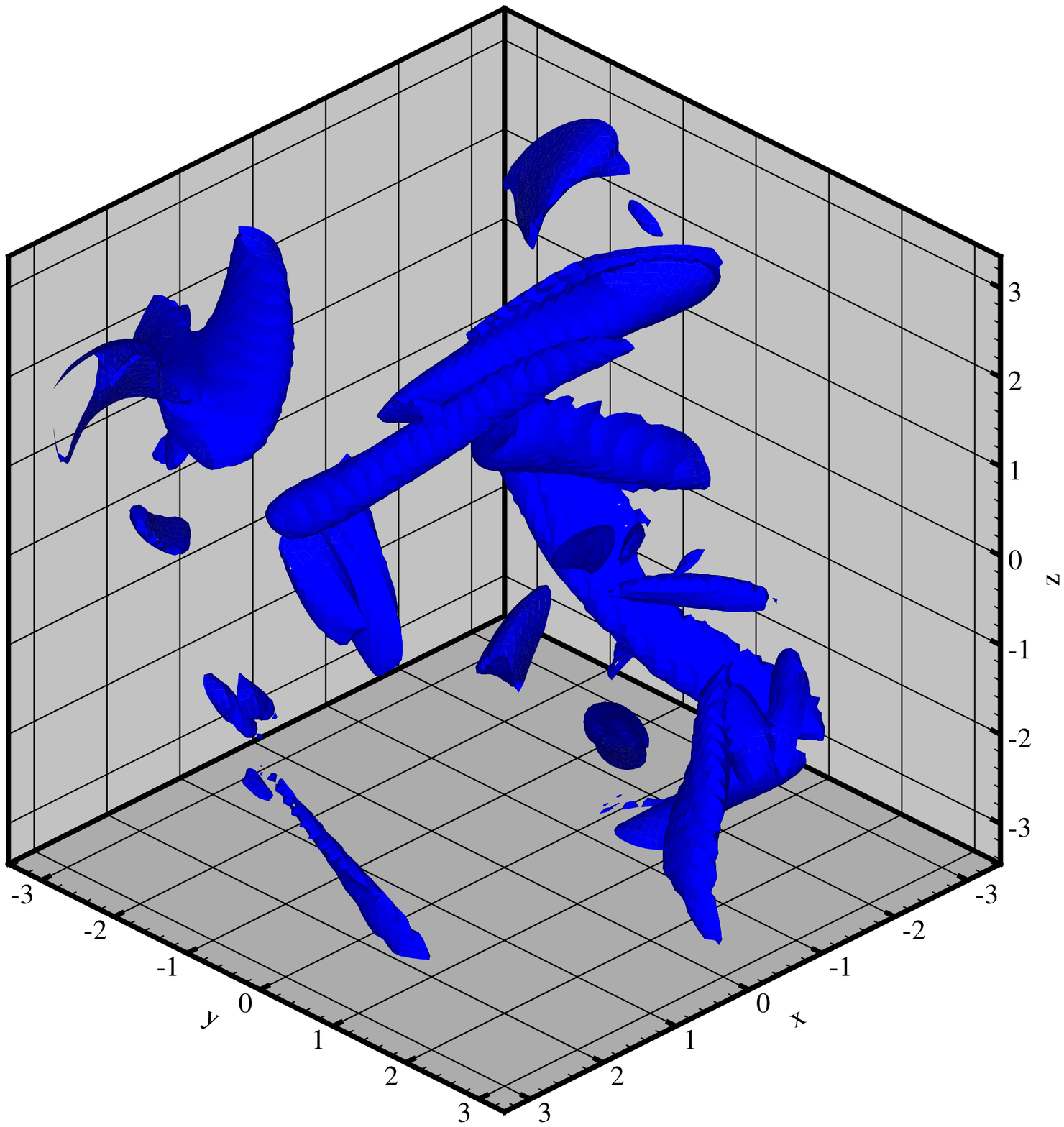,width=7.cm}}
   \caption{ Vortical structures for the Newtonian, $De=0$, (top) and viscoelastic case,
             $De=.38$, (bottom).  \label{F1} }
\efi
Table~\ref{Tab} presents the global parameters of a typical viscoelastic simulation,
$De=.38$, together with a corresponding Newtonian case.
In the table, the Taylor microscale is defined as 
$\lambda = (15/2 \, u_{rms}^2/\Omega)^{1/2}$, with $\Omega = 1/2 <\zeta^2>$, 
$\zeta$ being the vorticity. The average value of the free-energy is denoted by 
$\bar a$, while the other symbols have been already defined in the previous sections.
Concerning the Kolmogorov scale, $\eta$, for the viscoelastic 
case this quantity is not uniquely defined. 
Actually, for Newtonian turbulence one assumes that the small
scale dynamics is controlled by dissipation and viscosity, originating the classical
definition of the Kolmogorov length as $\eta = (\nu^3/\epsilon)^{1/4}$.
In the present case, several other parameters related to the more complex rheology
may in principle affect the small scale dynamics. In the table we have conventionally
assumed $\eta_T = (\nu_T^3/\epsilon_T)^{1/4}$.

\begin{table}
\begin{center}
\begin{tabular}{ccccccccccc}
$De$  & $Re_{\lambda}$ & $u_{rms}$ & $\Omega$ & $\bar a$ & $\bar{\epsilon}_T$ & $\bar{\epsilon}_N$ & $\bar{\epsilon}_P$ & $\lambda$ &
$\eta_T$   \\  \hline 
0     &   75           &  .806      &   10.6   &  -  &  .156        &   .156       &     -        &   .68     & .040     \\ 
.38   &   170          &  .9685      &    5.1   &  .09 &  .231        &   .067       &     .164     &   1.18    & .036
\end{tabular}
\end{center}
\caption{Global parameters for the Newtonian, $De=0$, and the viscoelastic, $De=.38$, simulation. \label{Tab} }
\end{table}

By discussing the data in table~\ref{Tab}, the macroscopic effects of viscoelasticity 
can be summarized as follows.  
For fixed forcing amplitude, i.e. for fixed nominal Reynolds number in identical
boxes and with same total viscosity, $\nu_T$, we observe a substantial increase of the
total dissipation. Clearly, in the steady state, this corresponds to an equivalent 
increase in the power extracted from the external forcing.
For the viscoelastic case, most of the dissipation, about $71 \%$, occurs in the polymers.
The $u_{rms}$ increases  of about $20 \%$, with respect to the 
Newtonian simulation. The increase of the energy content in the flow is even more apparent
if we consider that, in addition to kinetic energy, a non-negligible contribution arises
from the free energy $\bar a$. In terms of total energy, 
$\bar E = 3/2 \, u^2_{rms} + \bar a$, the increase with respect to Newtonian turbulence 
is of the order of $48 \%$. Clearly, the increase in the total dissipation for fixed total 
viscosity amounts to a reduction of the conventional Kolmogorov scale.

A further effect is observed when comparing the Taylor microscale for the two simulations.
The increase in turbulent kinetic energy of about $44 \%$ is accompanied by 
\bfi[b!]
   \centerline{
   \epsfig{figure=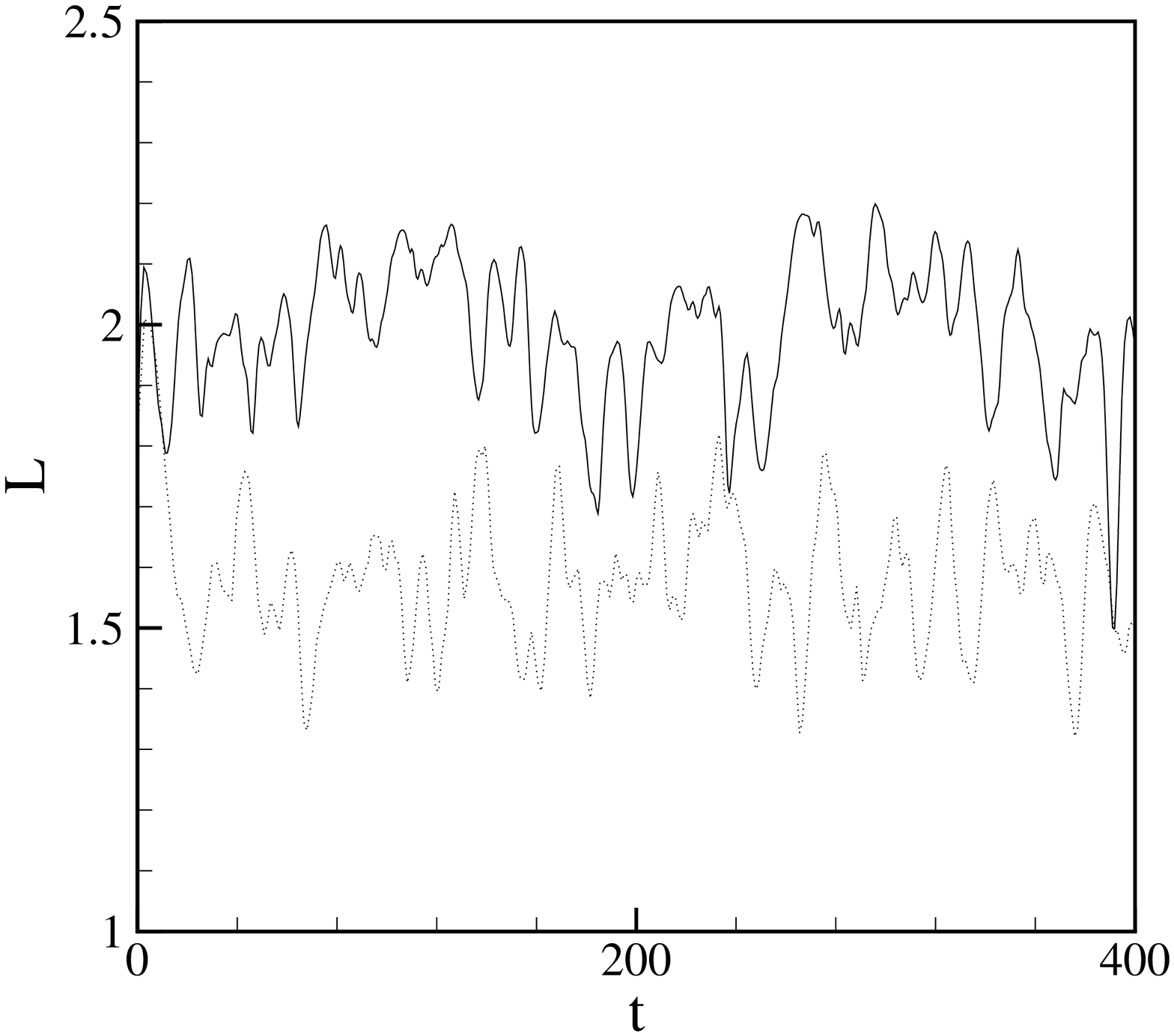,width=7.cm} }
   \centerline{
   \epsfig{figure=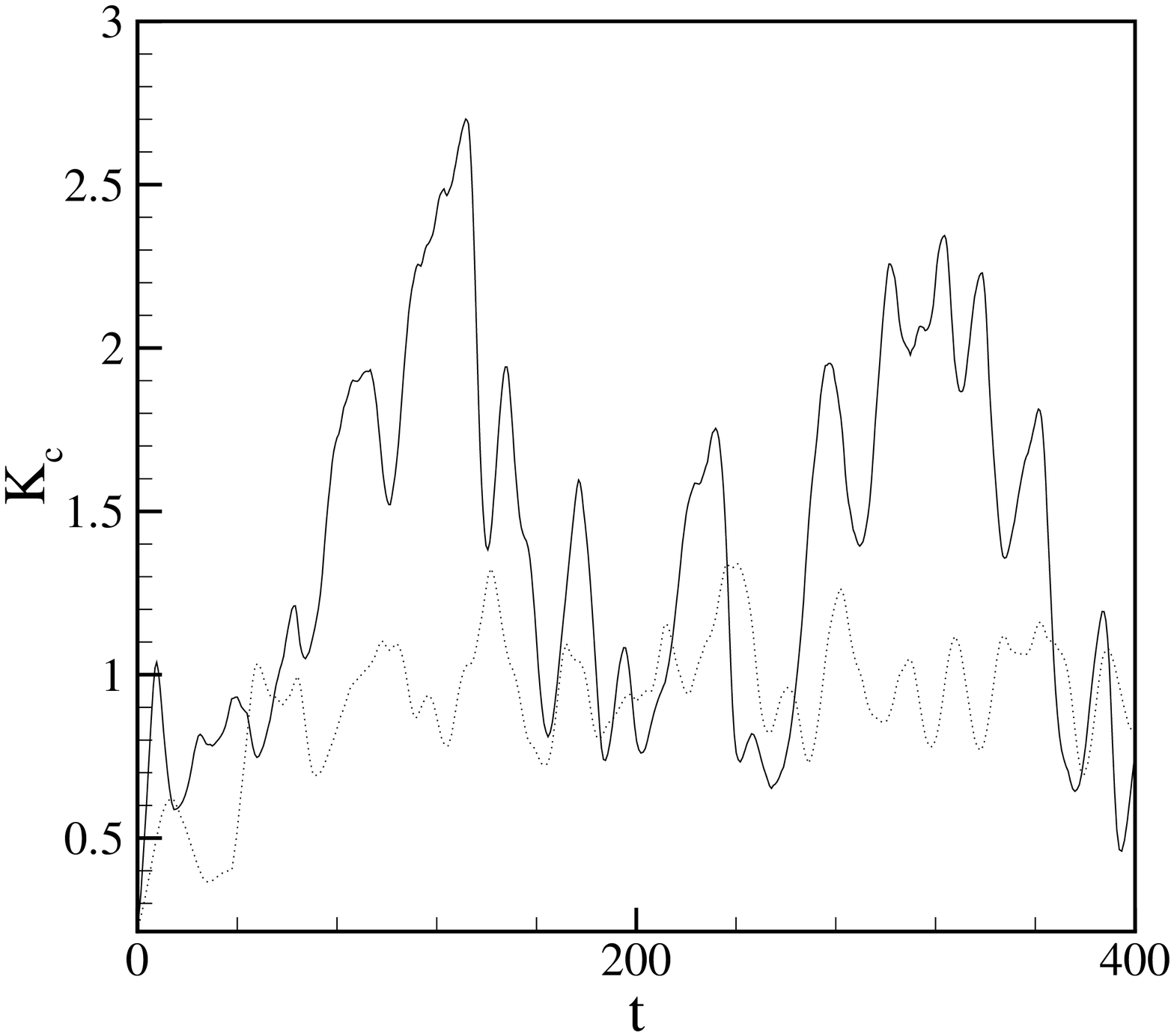,width=7.cm}}
   \caption{ Top: Time behavior of the integral scale for the Newtonian, (dotted) and viscoelastic, 
             (solid) case.  
             Bottom: Time behavior of the spatial average of the turbulent kinetic energy, 
             same symbols.
             \label{F2} }
\efi
\bfi[bh!]
   \centerline{
   \epsfig{figure=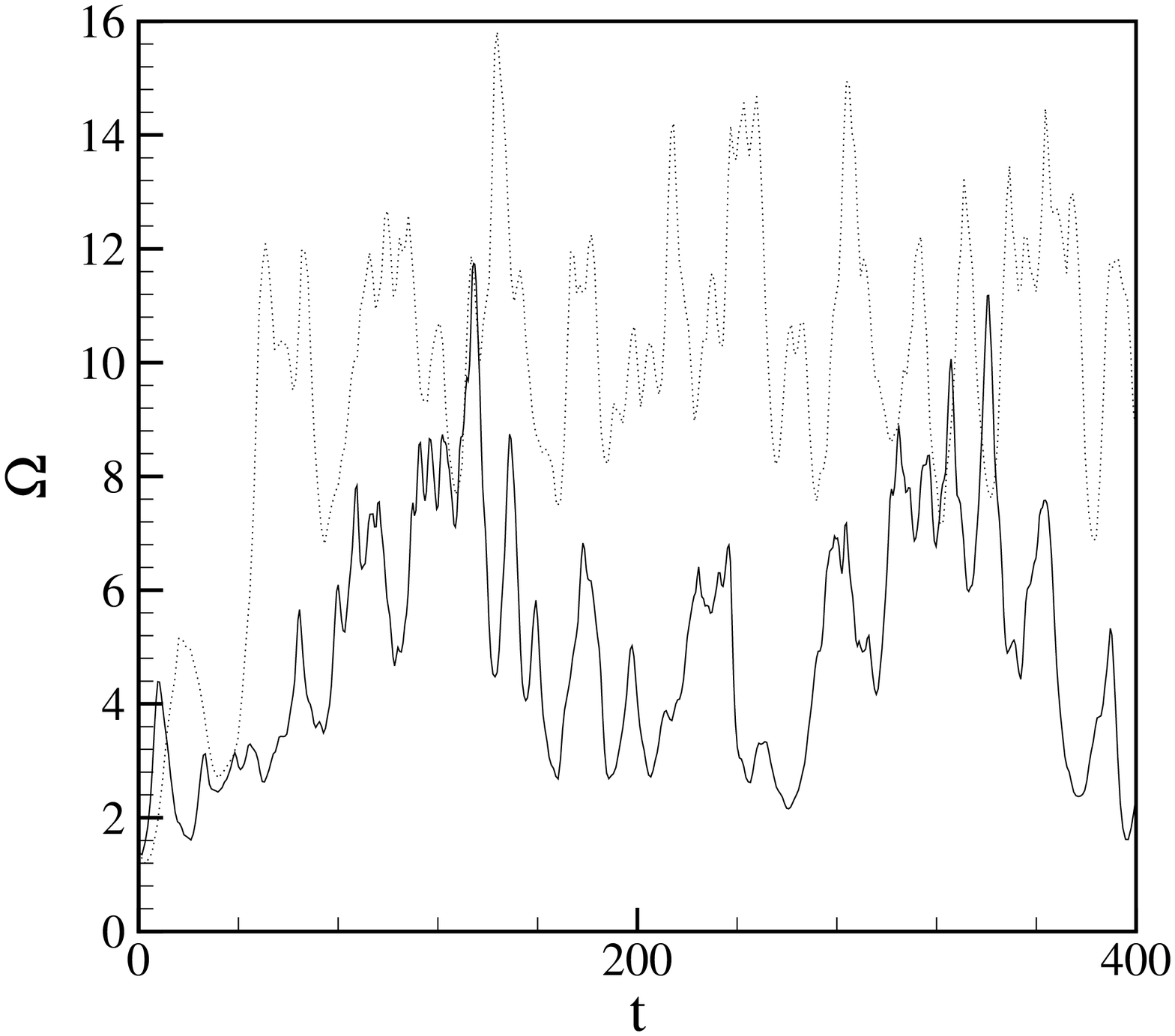,width=7.cm} }
   \centerline{
   \epsfig{figure=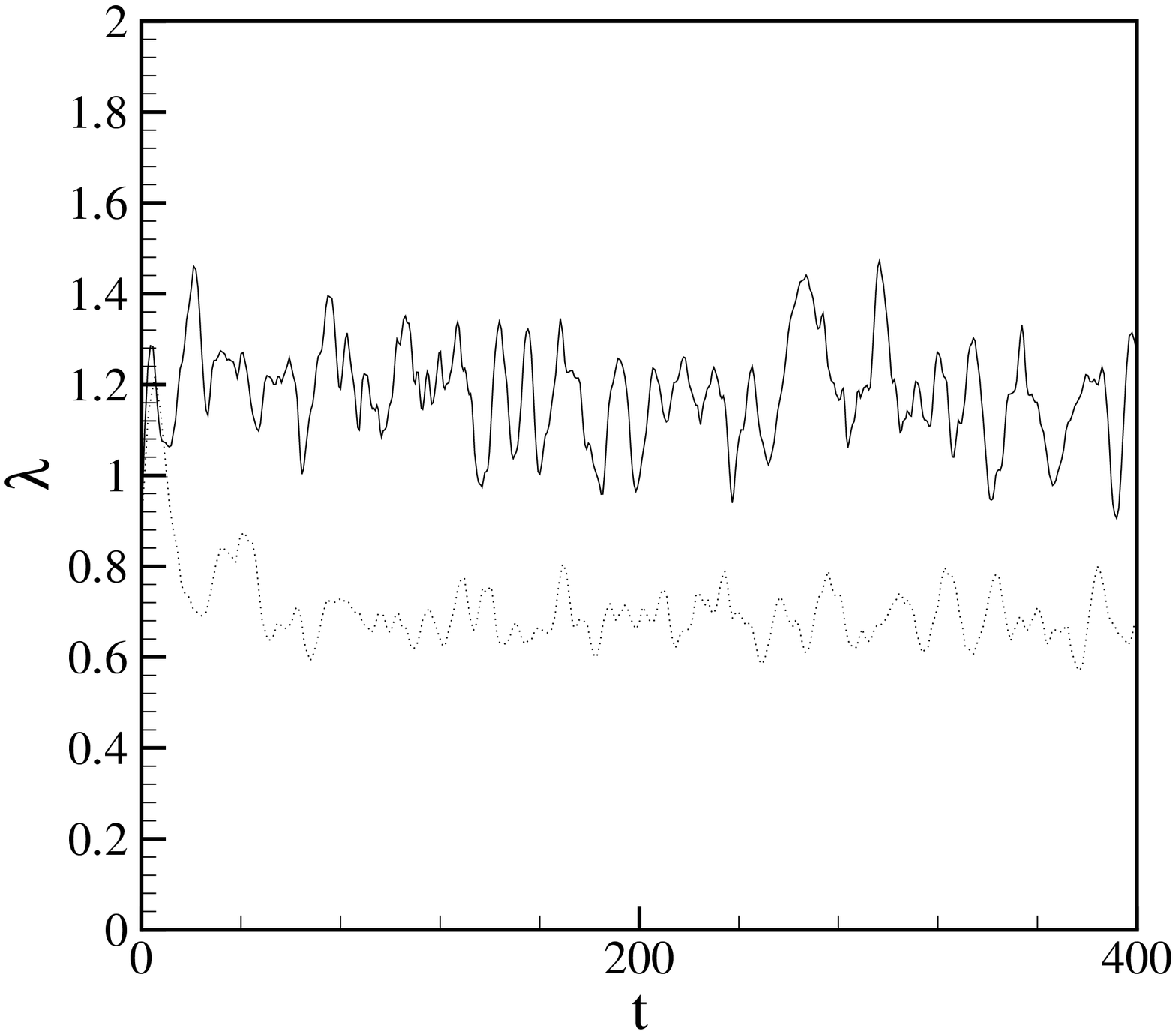,width=7.cm}}
   \caption{ Top: Time behavior of the spatial average of the enstrophy for the 
             Newtonian, (dotted), and  the
             viscoelastic case, (solid).  
             Bottom: Time behavior of the Taylor microscale, same symbols.
             \label{F3} }
\efi
a reduction
of the order of $50 \%$ in the enstrophy, giving an increase of $73 \%$ for the Taylor 
microscale. This suggests a substantial depletion in the energy content of the high 
wave-number modes. Consistently  a substantial increase in the Taylor-Reynolds number,
$Re_\lambda = u_{rms} \lambda/\nu_T$, is achieved.

From the numerical point of view, a comparable value for $Re_\lambda$ in Newtonian fluids
cannot be reached with only $64$ spatial modes. For the viscoelastic case, however, the 
requirements on the grid are less strict, since most of the dissipation, the polymeric 
contribution, is not associated to spatial gradients.
We observe in passing, that spectral calculation with the FENE-P model may be prone to
high-wavenumber instability, which is maintained under control by adding a small
amount of artificial viscosity in the algorithm for the conformation tensor
\cite{Beris}, see also \cite{Choi} for an alternative approach.

The evolution of the system, followed for about $50$ large-eddy turnover times, 
$T = (l_x/\hat f_0)^{1/2}  = 7.9$, is globally 
described in figure~\ref{F2}, where the time history of the instantaneous integral scale 
$L = \pi/2 \int_0^{K_{max}} 1/k \, {\hat e}(k) \, dk \, /\, u^2_{rms}$, where $\hat e$ is the
instantaneous energy spectrum, is shown on the top.
From the comparison between viscoelastic and Newtonian case, the average increase in $L$
is apparent. Clearly this is associated to the increase in the amount of energy 
in the energy containing range, and corresponds to the observation of larger structures
in the field. The presence of fewer but more energetic coherent structures is also
consistent with the existence of more pronounced intermittency cycles, more evident
in the history of the kinetic energy shown on the bottom part of the same figure.

Figure~\ref{F3} describes the evolution of enstrophy, top, and Taylor microscale, bottom.
On average, the enstrophy reduces  with respect to Newtonian turbulence, and presents large 
spikes following with slight delay the corresponding peaks found in the kinetic energy.
In figure~\ref{F3_bis} we report on top the time behavior of the spatial
average of the free energy.
Apparently, the amount of energy stored by the polymers is substantially
lower than the kinetic energy, already given in figure~\ref{F2} and reproduced 
\bfi[b!]
   \centerline{
   \epsfig{figure=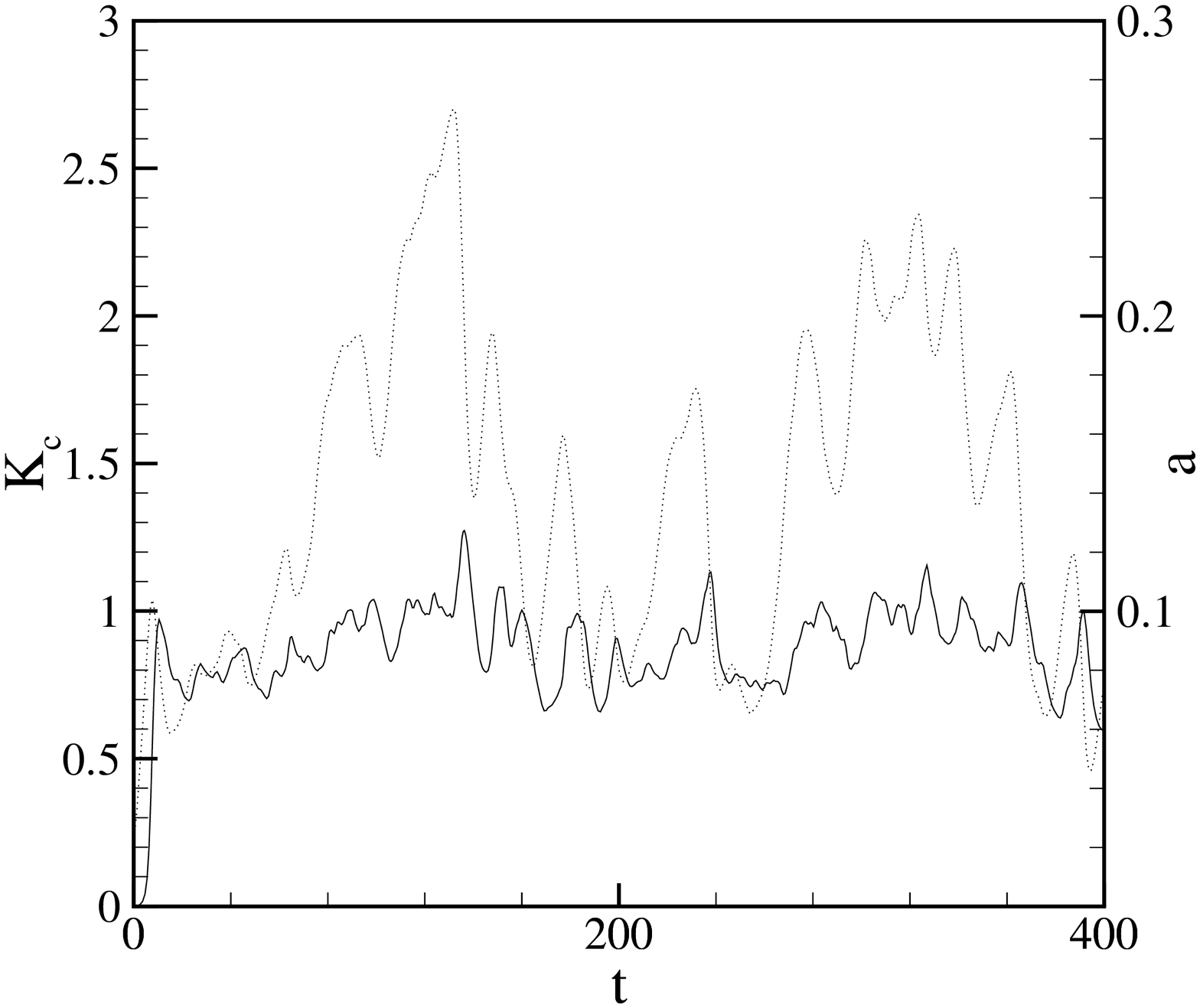,width=7.cm} }
   \centerline{
   \epsfig{figure=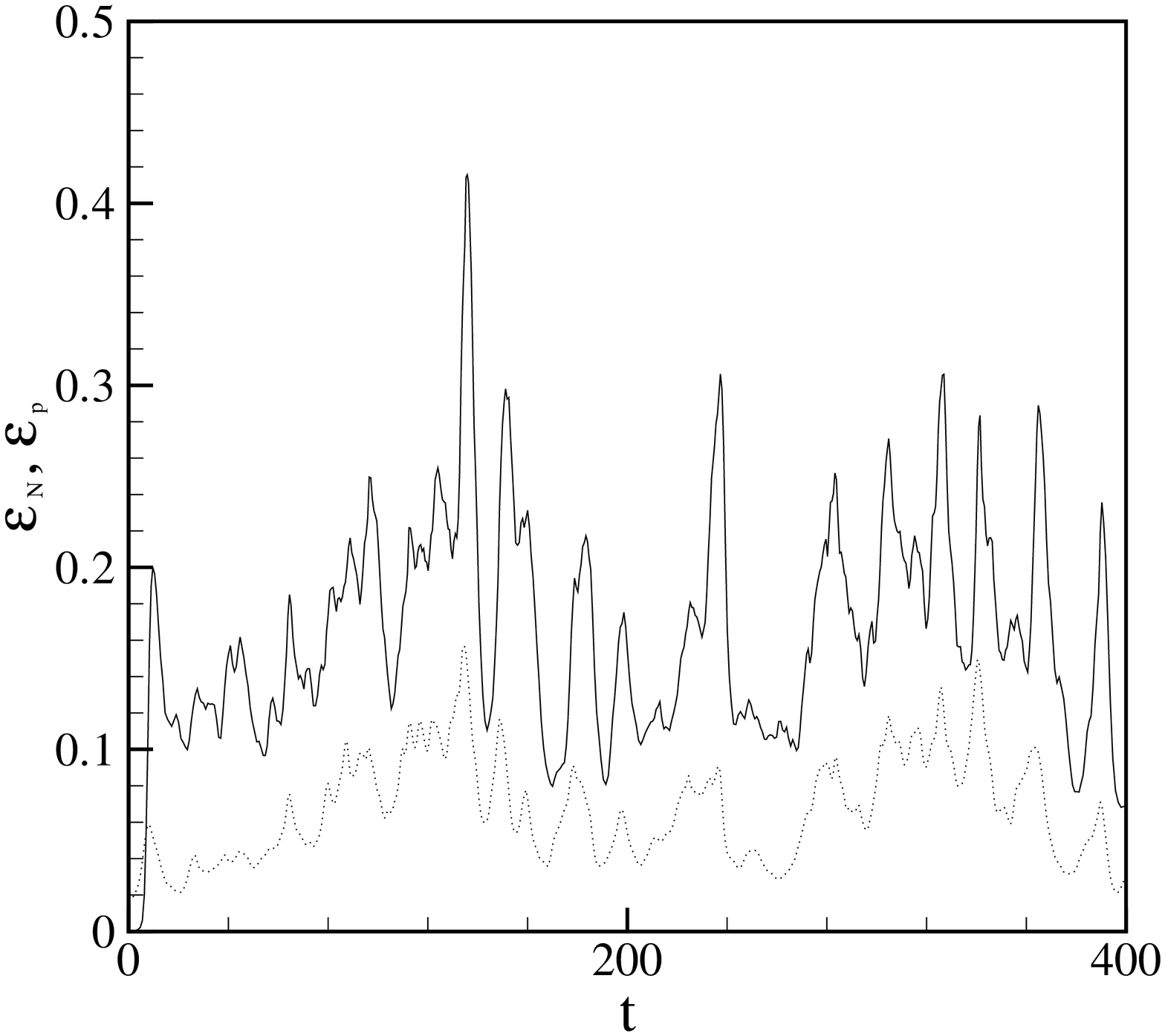,width=7.cm}}
   \caption{ Viscoelastic simulation, $De=.38$.
             Top: Time behavior of the spatial average of the free-energy, $a$, (solid)
             compared with the kinetic energy (dotted). Note the scales, differing by an order of
             magnitude, for free-energy on the right and kinetic energy on the left.
             Bottom: Time behavior of the spatial average of the Newtonian (dotted)
                     and the viscoelastic component (solid) of the dissipation.
             \label{F3_bis} }
\efi
here for convenience. 
From the comparison one may note the more irregular behavior of the free-energy with 
reduced peak to peak oscillations.
Finally, on the bottom of the figure we plot the evolution of the two forms of dissipation,
Newtonian and viscoelastic respectively. Here the trend is opposite, with a larger portion 
of dissipation occurring in the polymers.

\section{Scale by scale budget}

The Karman-Howarth equation (\ref{k_h_sphere}) provides the appropriate tool
to analyze the scale by scale budget for the velocity field.
To ease its interpretation one should keep in mind the term 

\[
- \frac{1}{\pi r^2} \frac{d}{dt} \int_{B_r}  C_{i i} dV_r \, ,
\]

\noi originally present at the left-hand side of the equation, that 
has been dropped due to the assumed stationarity of the field.

A negative flux through the boundary of $B_r$, 

\bea
\label{Phi_T}
\Phi(r) \, =  \, \frac{1}{4 \pi r^2}
\oint_{\partial B_r} <\delta V^2 \delta V_k + 2 \,
 T^*_{k i} \, \delta V_i> n_k \, d S_r   \, ,
\eea

\noi corresponding to a flux towards the inside of $\partial B_r$,
implies a negative contribution to the integral up to $r$ of the 
correlation, i.e. an average depletion of correlation in the scales smaller than $r$.
For stationary flows, the depletion balances the feeding operated
by the forcing, 

\[
 F(r) \, = \,   -4/3 \, <f_i \, u_i >  \, r   = -4/3 \, \bar{\epsilon}_T \, r  \, ,
\]

\noi consistently with a steady correlation at all scales.

For polymers, the flux is split into two parts, the classical
part which,  for isotropic conditions, can be expressed in terms of
the third order longitudinal structure function,

\bea
\label{Phi_C}
\Phi_c(r) \, = \, 
\frac{1}{4 \pi r^2}
\oint_{\partial B_r} <\delta V^2 \delta V_k > n_k \, d S_r   \, =\, 
\frac{3}{5} <\delta V^3_\parallel> \nonumber \\
\eea

\noi and the additional term associated with the viscoelastic reaction,

\bea
\label{Phi_P}
\Phi_p(r) \, = \, 
\frac{1}{4 \pi r^2}
\oint_{\partial B_r} <T^*_{k i} \, \delta V_i> n_k \, d S_r  \, .
\eea

Figure~\ref{F4} shows all the contributions appearing in the steady state
Karman-Howarth equation, and, in particular,
on the top, it reports the budget of the Newtonian simulation for comparison.
Considering the Newtonian case first,
the straight-line with negative slope corresponds to $-4/3 \bar{ \epsilon}_N r$,
filled squares. The positive correlation between forcing increments and
velocity increments is represented by the filled diamonds.
The nonlinear transfer term, $\Phi_c$, is given by the filled circles, while the 
viscous correction corresponds to the filled triangles.
All terms sum up to zero within the accuracy of the available statistics.
Clearly, the viscous correction is present at all scales, confirming that the Reynolds
number of the simulation is too small to achieve a proper inertial range.
A further point to be highlighted is the non-negligible contribution of the correlation
between forcing increments and velocity increments.
Nonetheless, for a fair range of scales, up to $r/\eta_T \approx 30$, the balance 
involves only the nonlinear transfer term , $\Phi_c$, and the forcing $F$. 
$\Phi_c$ displays a behavior in terms of separation which, a part from 
minor corrections due to the other terms, follows the linear trend of $F$.

The same kind of budget is discussed for the viscoelastic case in the bottom part of the figure. 
The filled symbols have the same meaning defined for the
Newtonian case, except that, now, the transfer term, filled circles,  corresponds to
the entire flux $\Phi$, i.e. it accounts also for the polymeric contribution.
As we may see,  the picture grossly reproduces that already described for Newtonian 
turbulence. 
One should note that the forcing term, $F$, now
involves the entire dissipation $\bar{\epsilon}_T$, i.e. the sum of the 
\bfi[b!]
   \centerline{
   \epsfig{figure=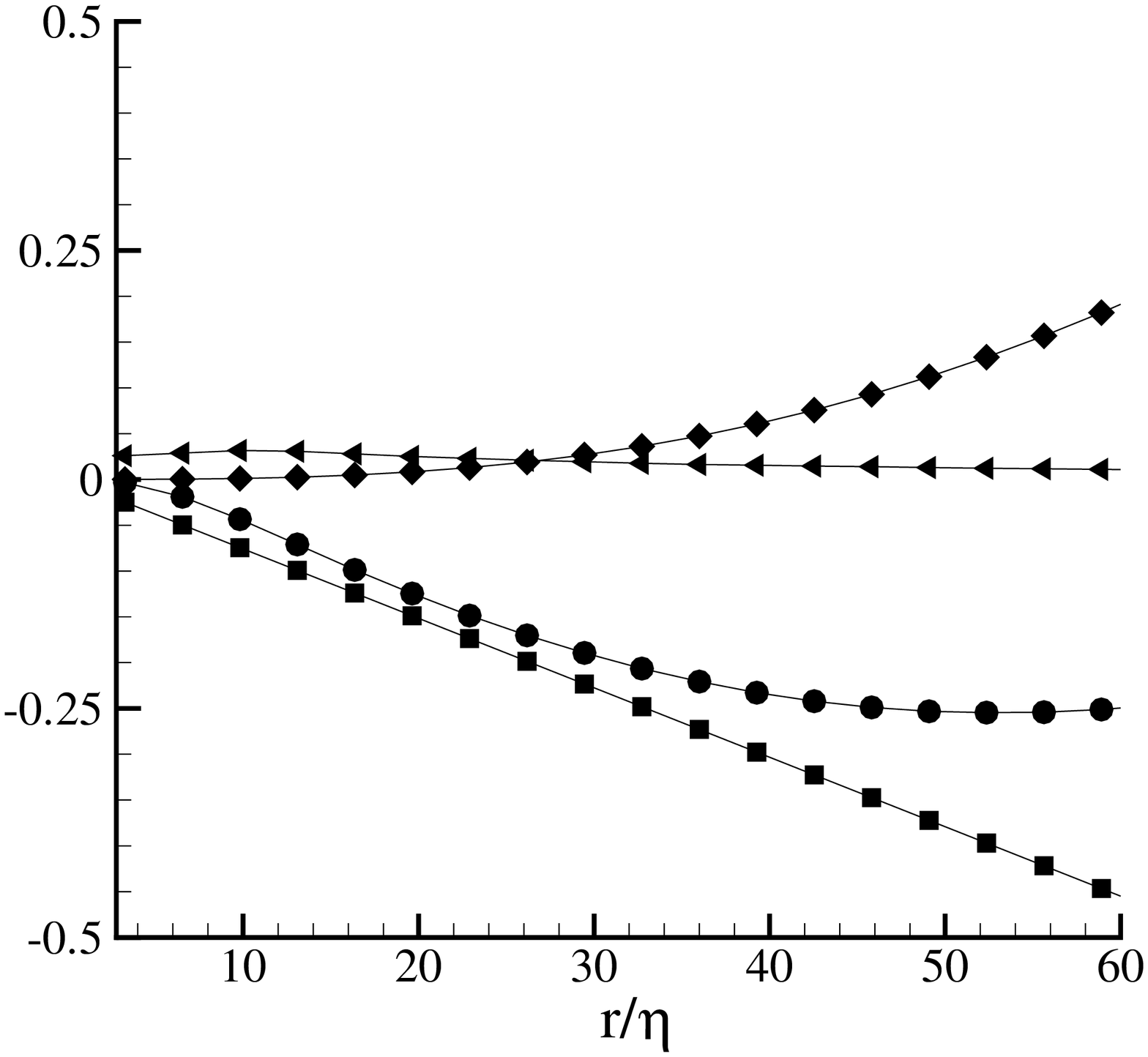,width=7.cm} }
   \centerline{
   \epsfig{figure=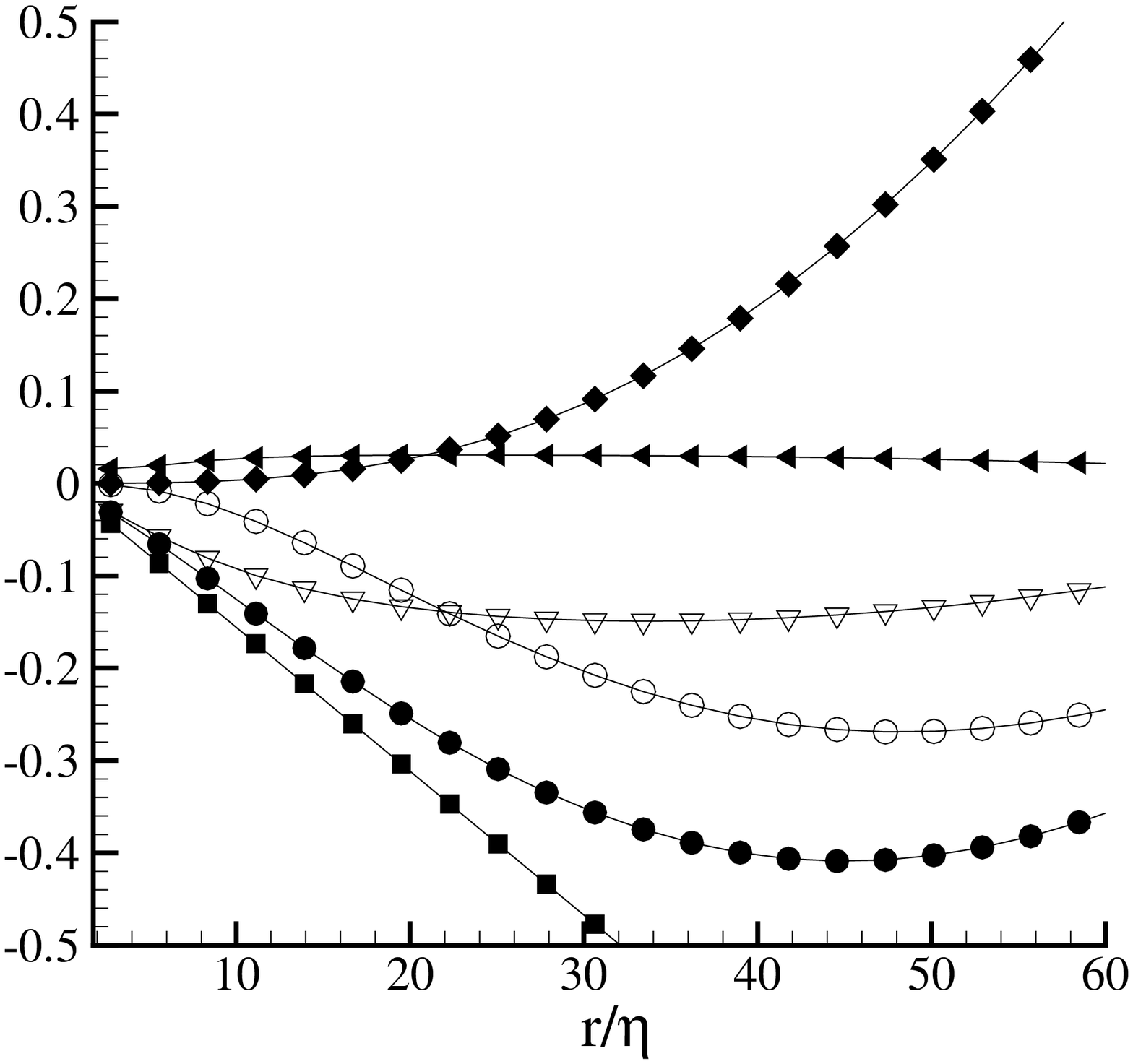,width=7.cm}}
   \caption{ Scale by scale budget for the Newtonian (top) and the viscoelastic case 
             (bottom). Symbols are defined in the text. \label{F4} }
\efi

Newtonian and the polymeric contributions. 
The slope is then substantially larger than in the Newtonian
case displayed on the top. Not unexpectedly, the  global flux term is qualitatively
similar to the corresponding Newtonian one, with a substantial range of scales where
the linear trend defined by  $F$ is reasonably well reproduced by $\Phi$.
This emerges quite clearly from the plots in figure~\ref{F5} where the budgets 
for Newtonian and viscoelastic turbulence are superimposed the one to the other, after a 
rescaling proportional to the respective values of the global dissipation.
Also for the polymers, we are in presence of a direct cascade, with a negative 
global flux.

Returning to the plots on the bottom of figure~\ref{F4},
the most interesting feature is represented by the open symbols, circles and squares, which 
give the splitting of the flux into the classical contribution associated to the
third order velocity structure function, $\Phi_c$ (circles), and to the polymer 
contribution, $\Phi_p$ (triangles), respectively.
The larger scales are less affected by the polymers, with $\Phi_p$
becoming sub-leading with respect to $\Phi_c$.
One may conjecture that, at sufficiently large scale and for  Reynolds numbers large 
enough, the turbulence scalings are substantially unaffected by the polymers.
Under this assumption, we see that the total dissipation $\bar{\epsilon}_T$ may be
taken to measure a purely inertial energy transfer at large
scales. This attaches to our conventional
definition of Kolmogorov scale the obvious meaning of the length-scale where dissipation
would occur in the absence of polymers.

Within the scenario according to which the large scales are unaffected by the polymers,
the time criterion of Lumley \cite{Lumley2} fixes the scale $r^*$ below 
which the polymers begin to feel the turbulent fluctuations. 
This happens when the eddy-turnover time of classical turbulent eddies, i.e. without 
the effect of the viscoelastic reaction, equals the principal relaxation time of the polymer 
chains. Assuming Kolmogorov scaling for the velocity increments, this leads to

\bea
\label{TCS}
r^* & = & (\bar{\epsilon}_T \, \tau^3)^{1/2}
\eea

\noi or, scaled with the conventional Kolmogorov length, to

\bea
\label{TCS_K}
r^*/\eta_T & = & \tau^{3/2} \, (\bar{\epsilon}_T /\nu_T)^{3/4} \, .
\eea

\noi From the data given in table~\ref{Tab}, for our simulation we have 
$r^*/\eta_T \simeq 70$,
within a computational box of $l_x/\eta_T \simeq 174$, suggesting that the forcing,
applied to the first shell of wavenumbers,
acts just above the scale where the polymers already feel the turbulent fluctuations.
In fact, a better way to apply the time criterion is to identify the eddy-turnover
time as function of separation from the actual data provided by the DNS. 
This is done in figure~\ref{F5_bis}, where as characteristic time for the  scale $r$
we assume $r/\Phi_c^{1/3}$. From the figure we infer that our viscoelastic turbulence
achieve an eddy-turnover time comparable with the viscoelastic relaxation time,
horizontal line in the figure, at a scale considerably smaller then predicted on purely
dimensional grounds, leaving a reasonable range of between the forcing and Lumley scale.

\bfi[t!]
   \centerline{
   \epsfig{figure=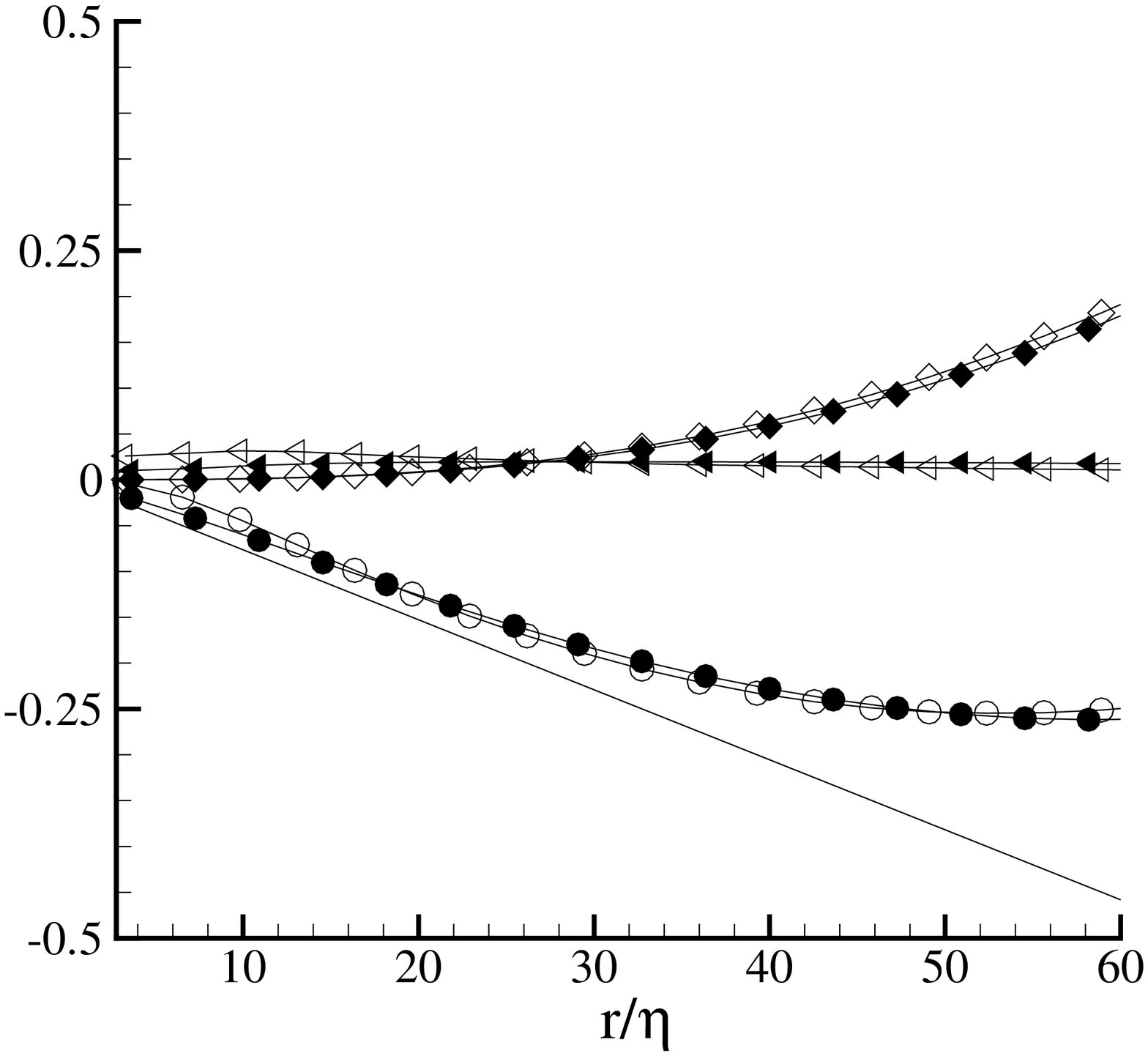,width=7.cm}  }
   \caption{Re-scaled budget for the Newtonian (open symbols) 
            and the viscoelastic (filled symbols).
             \label{F5} }
\efi

Coming back to our comparison between the two components of the flux, at small scales 
we find that the component originated by the polymers takes over, 
entailing the reduction of the third-order structure function.
The two curves, open circles and squares, clearly identify a cross-over scale
$\ell_p$ according to the equation

\bea
\label{CO}
\Phi_p(\ell_p) \, = \, \Phi_c(\ell_p) \, .
\eea

\noi For our case the balance is achieved at $r/\eta_T \simeq 22$, while 
$\lambda/\eta_T = 33$.
Observe for comparison that, in the corresponding Newtonian case, $\lambda/\eta$ is 
about half the value.
By inspection of the figure, we also find that all along the available range of scales
the polymers reaction is dynamically relevant for the turbulent fluctuations, 
with $\Phi_p$ always comparable in magnitude with $\Phi_c$.

\noi Dimensionally, the right-hand side of eq.~(\ref{CO}) can be estimated as 
$\bar{ \epsilon}_T \, \ell_p$, as follows by roughly assuming a purely classical inertial 
scaling for $r \ge \ell_p$.
On the other hand, $\Phi_p(\ell_p)$ should be somehow proportional to 
$\sigma^{**} \, \nu_p/\tau \, \bar{\epsilon}_T^{1/3}\, \ell_p^{1/3}$, where, again,
Kolmogorov scaling is used to estimate $\delta V$ in eq.~(\ref{Phi_P}).
Accounting for the constitutive relation (\ref{ex-stress}), $T^*$ is expressed
in terms of the, as yet, undetermined factor $\sigma^{**}$ which should account for 
the level of stretching achieved in the polymers at the cross-over scale. 
This yields the expression

\[
\ell_p = \frac{(\sigma^{**} \, \nu_p/\tau)^{3/2}} {\bar{\epsilon}_T} \, ,
\]

\noi which can be used to build the dimensionless quantity

\bea
\label{N-N}
\sigma^{**} & = & \tau/\nu_p \, (\ell_p \bar{\epsilon}_T)^{2/3} \, .
\eea

For the present case we have $ {(\nu_p/\tau)^{3/2}}/{\bar{\epsilon}_T} \simeq 1.4 \, 10^{-5}$,
which, after considering that $\ell_p/\eta_p \simeq 22$, 
\bfi[t!]
   \centerline{
   \epsfig{figure=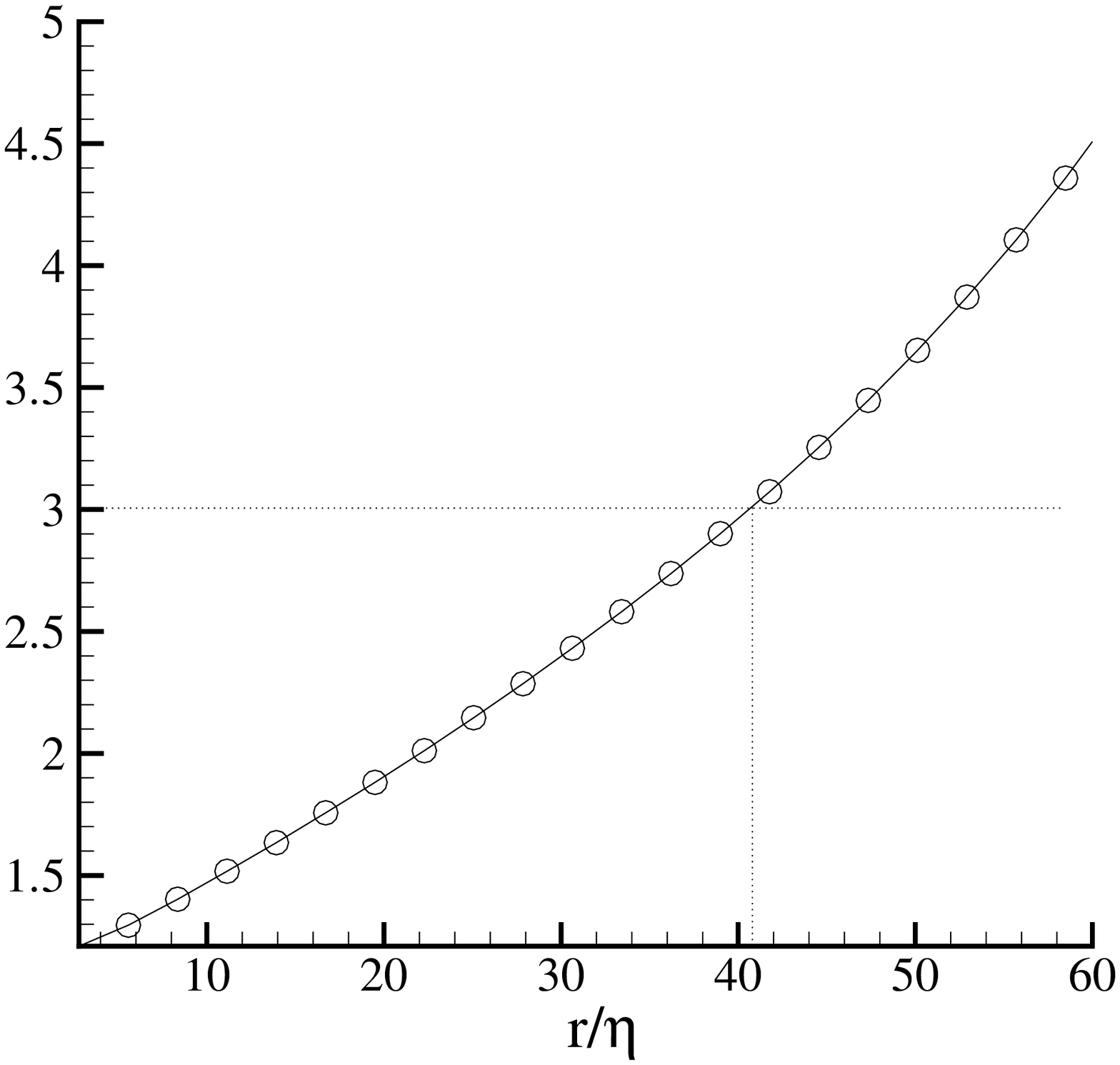,width=7.cm} 
                                                  }
   \caption{Eddy-turnover time, $r/\Phi_c^{1/3}$, vs scale $r$. The horizontal
            line corresponds to the relaxation time, $\tau$.
             \label{F5_bis} }
\efi
yields  $\sigma^{**} \simeq 1400$.
The parameter $\sigma^{**}$ allows to quantify the effective stretching of the chains at 
cross-over by assuming  $\sigma^{**} \propto f^{**} R^{**}_{kk}/\rho^2_0$, 
see eq.~(\ref{ex-stress}).

The quantity 

\bea
\label{sigma}
\sigma = f(R_{kk}; \rho_{max}, \rho_0) R_{kk}/\rho^2_0
\eea

\noi is plotted versus the elongation $R_{kk}/\rho^2_0$ on the right of figure~\ref{F5_tris}.
We see clearly the finite extensibility effect, with the unbound growth of $\sigma$ 
as the critical extension $\rho^2_{max}/\rho^2_0$ is approached from the left.
Note that, at small elongation, $\sigma$ goes almost linearly with $R_{kk}/\rho^2_0$.
The horizontal line plotted in the figure corresponds to the value $\sigma^{**}$
we have estimated from the numerical simulation. The intercept of this line
with the curve representing $\sigma$ lays relatively  far from the critical extension,
though well within the region of non-linear elastic response.

It may be instructive to compare the elongation associated to the cross-over scale
to the pdf of $R_{kk}/\rho^2_0$.
\bfi[t!]
   \centerline{
   \epsfig{figure=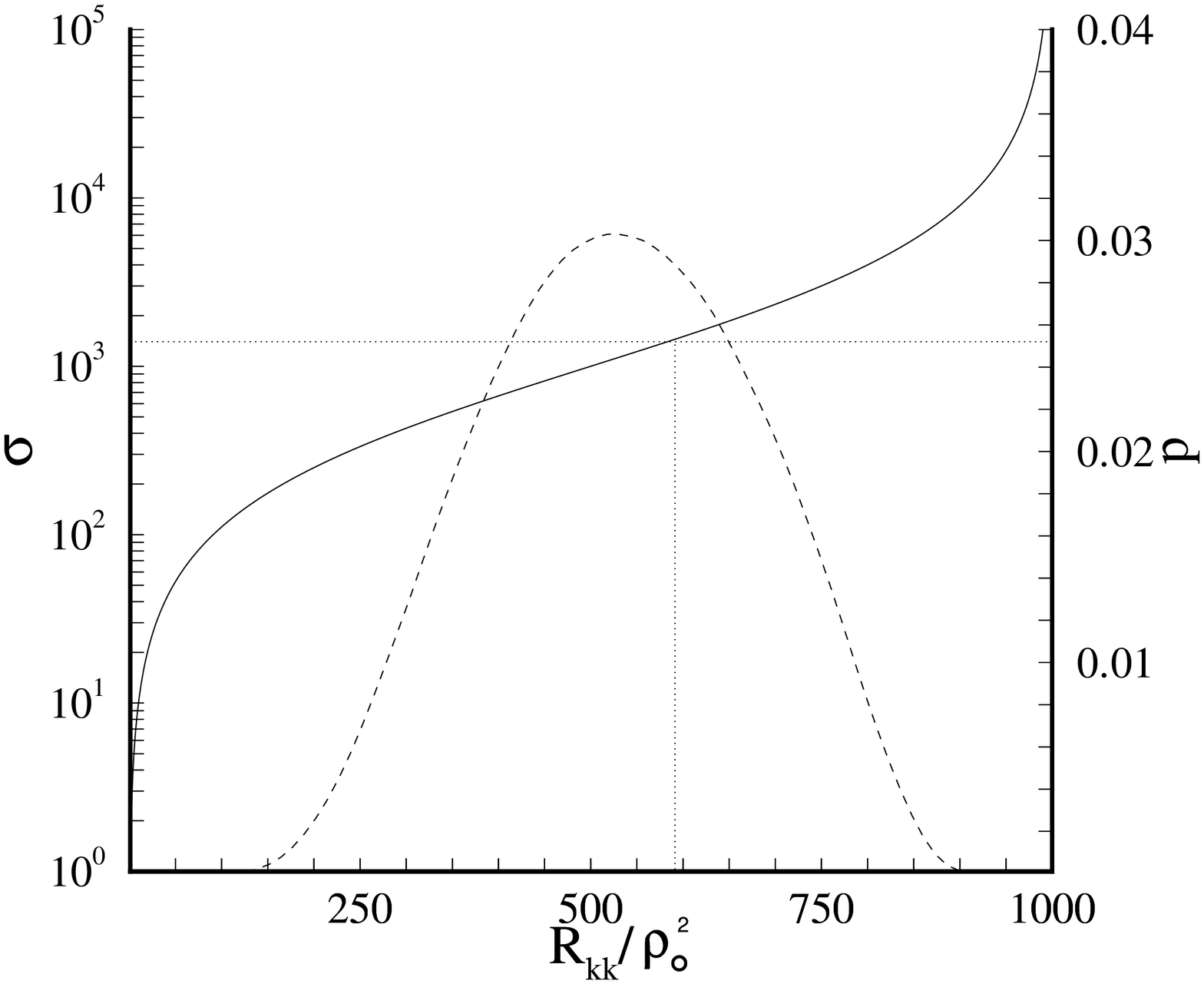,width=7.cm} }
   \caption{$\sigma$, see eq.~(\ref{sigma}), vs. polymers elongation, $R_{kk}/\rho_0^2$.
             \label{F5_tris} }
\efi
To this purpose, we superimpose to the plot of $\sigma$ given in figure~\ref{F5_tris} the 
histogram of the elongation as obtained by the numerical simulation. 
The effective elongation at cross-over, denoted by the vertical line, lies to
the right of the maximum of the histogram, indicating a level of stretching 
substantially larger then either the most probable or the average elongation.
Clearly the estimate here proposed should be understood as an order of magnitude
analysis, to allow the comparison of different flow conditions.

To complete the analysis, in figure~\ref{F6} we address the Yaglom-like equation for 
the free-energy.
As seen in the top part of the figure, the left and the right hand side of 
eq.~(\ref{Y_sphere}) balance within the accuracy of the available statistics. 

The fluctuations of excess-power are converted into a free-energy flux with a negative
sign, as wee see from the figure. Hence, also for the microstructure, we are in 
presence of a direct cascade. We like to stress that 
the direction of the cascade 
is not a priori obvious in this case, since we miss the standard term, linear in 
the 
\bfi[t!]
   \centerline{
   \epsfig{figure=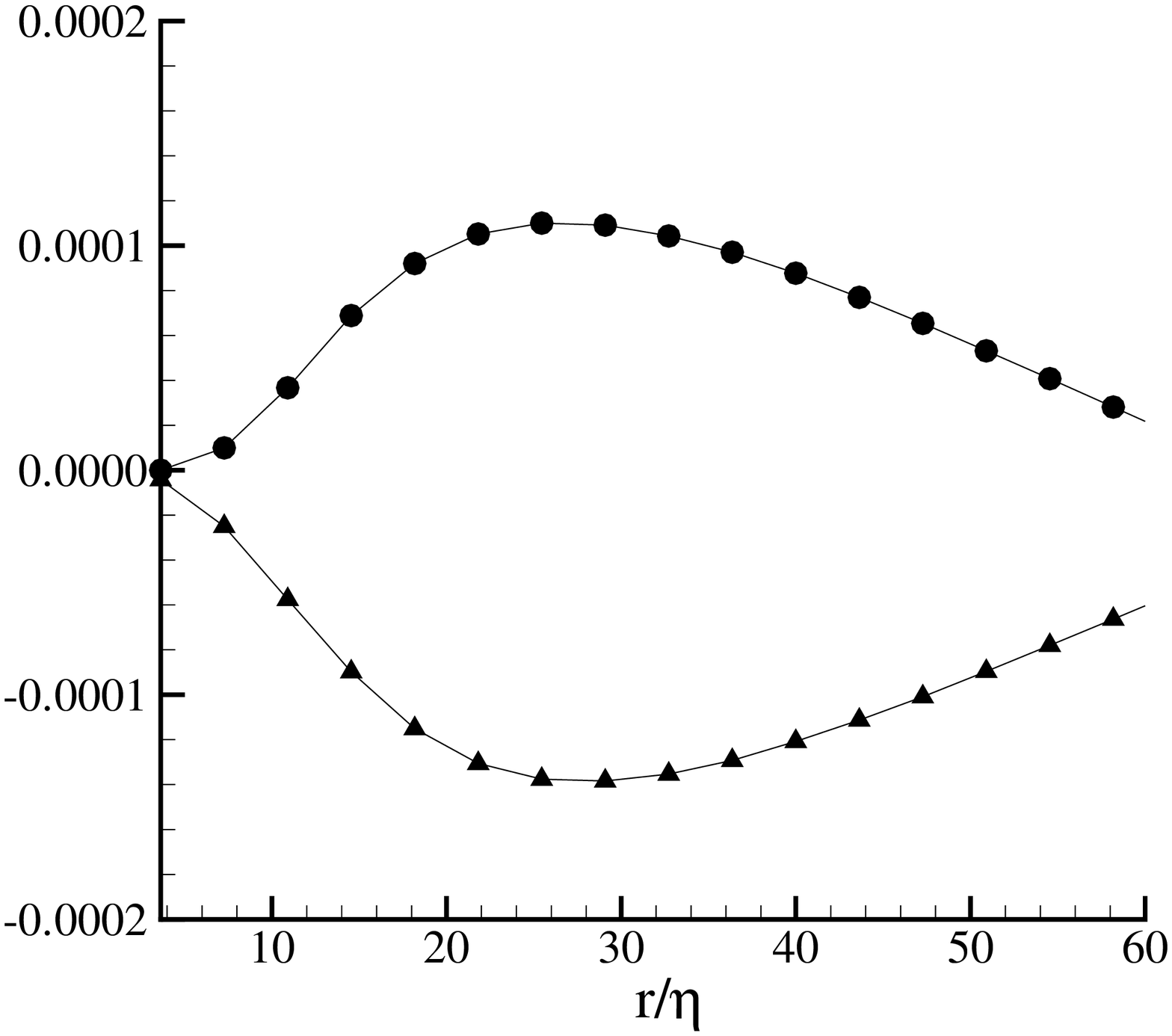,width=7.cm} }
   \centerline{
   \epsfig{figure=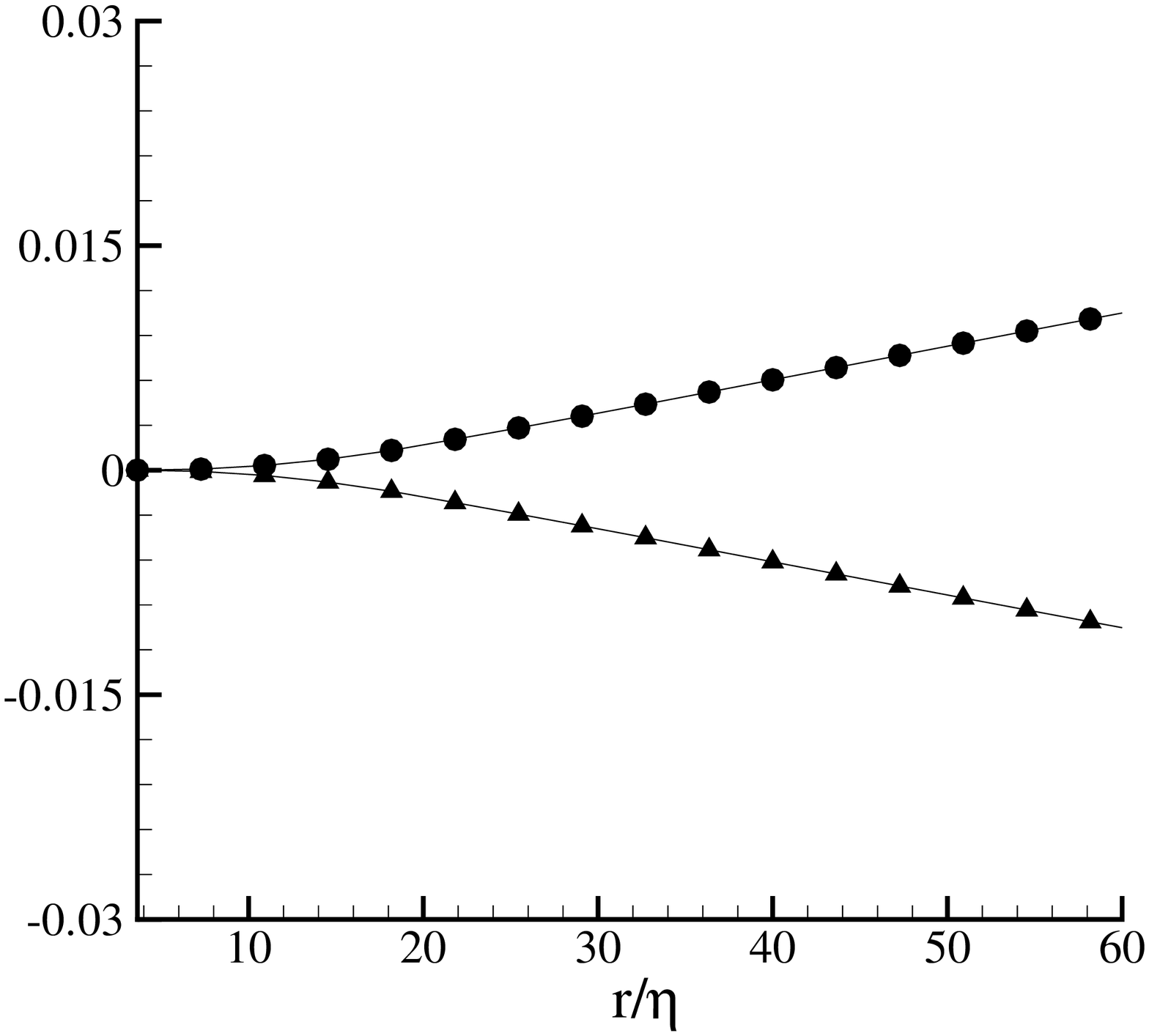,width=7.cm}}
   \caption{Top: The free-energy flux (filled triangles), and the forcing term given by the
            excess-power (filled circles) according to Yaglom equation.
            Bottom: The two contributions to the excess-power term in the Yaglom equation,
                   stress-power term (filled circles) and polymeric dissipation term 
                   (filled triangles).
             \label{F6} }
\efi
separation and related to the dissipation of the scalar, which usually 
determines the sign of the energy flux.

From the figure, the free-energy flux,

\[
\Phi_a \, = \, <\delta a^2 \, \delta V_\parallel> \, ,
\]

\noi manifests a pronounced minimum in correspondence with the cross-over scale $\ell_p$, 
suggesting that, near cross-over the activity of the polymers is particularly intense
and the polymers tend to become less active as the scale is increased.
The bottom part of the figure gives the decomposition of the right-hand side of 
equation~(\ref{Y_sphere}), with the positive term 

associated to  the stress power
denoted by filled squares and that associated with the polymeric dissipation
indicated by the filled triangles.

Order of magnitude  considerations confirm that the data shown in the bottom part of
figures~\ref{F4} and in figure~\ref{F6} are consistent.
At cross-over, $r=\ell_p$, we find from figure~\ref{F4} that 
$\Phi^{**}_c \simeq <\delta {V^{**}}^3> \simeq .12$, suggesting $\delta V^{**} \simeq .5$.
From figure~\ref{F6} we also have $<\delta {a^{**}}^2 \delta V^{**}> \simeq .00015$.
It follows $\delta a^{**} \simeq .017$ a value considerably close to 
$\delta {V^{**}}^2 \simeq .025$.
In other words, at the cross-over scale fluctuations of free-energy, or more loosely of
elastic energy, and fluctuation of turbulent kinetic energy become comparable,
as it should be expected.

\section{Concluding remarks}

The interaction between long chain polymers in dilute solution and turbulence
is known to have a profound influence on the structure of turbulence.
Several possible explanations have been proposed, with the eventual aim to  
understand the drag reducing properties manifested by such fluids in wall
bounded turbulent flows. Even the most powerful approach, however,
has to face the elusive mechanism by which polymers dynamics and turbulence
are mutually coupled. Among others, specially noteworthy is the theory
of De~Gennes and Tabor, who proposed a cascade model for the viscoelastic 
turbulence of dilute polymers solutions. Their elegant theory issues from
considering homogeneous isotropic conditions and rests upon
suitable assumptions on the self-similar stretching of polymers chains
under a fluctuating velocity field.
The entire model is essentially phenomenological in nature, and should thus 
be contrasted with experimental observations. 

In this context, we believe to have provided here a reasonable framework 
based on two essential ingredients. 
The first ingredient is the direct numerical simulation of the natural flow
where the cascade issue should be addressed, namely homogeneous isotropic turbulence 
numerically modeled as a triply-periodic box with random forcing at large scales.
Beside the obvious implementation in terms of spectral algorithms, this step
required the selection of a good rheological model for the polymers.
Based on previous experience, our natural choice has been the FENE-P model,
known to capture reasonably well the drag-reducing effects still maintaining
an affordable level of computational complexity.

The second ingredient is the extension to viscoelastic fluids of two of the, 
let's say, few exact results known in classical turbulence theory, namely
the Karman-Howarth equation for the velocity and the Yaglom equation for scalars.
As a characteristic feature of viscoelastic turbulence, the coupling between
velocity and extra-stress in the Karman-Howarth equation prevents recasting 
the result in the familiar form involving only longitudinal structure 
functions. Hence we have adopted a natural generalization in terms of suitable
surface and volume integrals in the space of separations.
Concerning the microstructure, the descriptor field is a second order 
tensor, making the derivation of the appropriate equation exceedingly cumbersome. 
We decided then to fucus our attention on the single most important
scalar quantity, namely the free-energy, whose Yaglom-like equation follows
straightforwardly.

The scale by scale budget based on the Karman-Howarth and Yaglom equation
has been used to understand the respective role of the traditional energy transfer
term versus polymeric transfer. We find a direct cascade occurring both in the kinetic
field and in the microstructure, for which, as already discussed, the sign of the energy 
flux is not a priori obvious.

The budget has allowed us to identify a cross-over
scale below which the polymers contribution becomes dominant, being sub-leading
with respect to inertial transfer at larger scales. 
This is the analogous of the scale defining the elastic limit in the theory of 
De~Gennes  and Tabor. It is here derived in a rigorous context, and evaluated on the
basis of a direct numerical simulation of a drag-reducing visco-elastic material
as described by the FENE-P model.

We find that the polymers substantially deplete the energy content of the
smallest scales, in agreement with recent experimental results, but they
affect the dynamics of the fluctuations at all the scales of our simulation.
The latter effect could be interpreted as  an artifact of the limited extension
of our computational domain which forced us to apply the external excitation 
relatively close to the scale defined by Lumley's time criterion. 
Nonetheless the present results make questionable the existence of 
a purely passive range, where polymers are deformed without back-reaction on 
the velocity field.

Based on the balance of terms appearing in the exact form of the Karman-Howarth 
equation we have proposed a dimensionless parameter which may give a quantitative 
measure of the stretching the polymers experience at cross-over, when elastic
effects become comparable with the corresponding kinetic energy.

Finally, a substantial level of intermittency is observed in the system, with
fluctuations in turbulent kinetic energy considerably larger than usually
observed in homogeneous isotropic turbulence of Newtonian fluids.
This effect is presumably associated with the existence of the additional 
mechanism of energy removal from the scales of the forcing.
In particular, a substantial part of the energy income does not follow the classical
route cascading towards viscous dissipation.
Instead, it is moved to the microstructure, to feed an additional cascade process
seemingly requiring a larger level of fluctuations.


\end{document}